\documentclass[useAMS,usenatbib]{mnras}

\usepackage{amsmath}
\usepackage{graphicx}
\usepackage{float}
\usepackage{gensymb}
\usepackage{amssymb}
\usepackage{tabularx}
\usepackage{subfig}
\usepackage{url}
\usepackage{float}
\usepackage{lineno}
\usepackage{multirow}
\usepackage{color}

\hypersetup{draft}

\DeclareRobustCommand{\ion}[2]{%
\relax\ifmmode
\ifx\testbx\f@series
{\mathbf{#1\,\mathsc{#2}}}\else
{\mathrm{#1\,\mathsc{#2}}}\fi
\else\textup{#1\,{\mdseries\textsc{#2}}}%
\fi}

\newcommand{\SFNUM}{$325 \:$} 
\newcommand{\ISONUM}{$158 \:$} 
\newcommand{\GNUM}{$167 \:$} 
\newcommand{\ewha}{{\rm EW}_{{\rm H\alpha}}}

\title[Quenching in groups]{The SAMI Galaxy Survey: Observing the environmental quenching of star formation in GAMA groups}
\author[Schaefer et al.]
{\parbox{\textwidth} 
{A.~L.~Schaefer$^{1,2,3,4}$,
S.~M.~Croom$^{1,3,5}$,
N.~Scott$^{1,3,5}$,
S.~Brough$^{3,5,6}$,
J.~T.~Allen$^{1,7}$,
K.~Bekki$^{8}$,
J.~Bland-Hawthorn$^{1,5}$,
J.~V.~Bloom$^{1,3}$,
J.~J.~Bryant$^{1,2,3,5}$,
L.~Cortese$^{8}$,
L.~J.~M.~Davies$^{8}$,
C.~Federrath$^{9}$,
L.~M.~R.~Fogarty$^{1}$,
A.~W.~Green$^{7}$,
B.~Groves$^{5,9}$,
A.~M.~Hopkins$^{7}$,
I.~S.~Konstantopoulos$^{7,10}$,
A.~R.~L\'opez-S\'anchez$^{7,11}$,
J.~S.~Lawrence$^{7}$,
R.~E.~McElroy$^{12}$,
A.~M.~Medling$^{9,13}$,
M.~S.~Owers$^{11}$,
M.~B.~Pracy$^{1}$,
S.~N.~Richards$^{14}$,
A.~S.~G.~Robotham$^{15}$,
J.~van~de~Sande$^{1,5}$,
C.~Tonini$^{16}$,
S.~K.~Yi$^{17}$
}
\vspace{0.4cm} \\
\parbox{\textwidth}{$^{1}$Sydney Institute for Astronomy, School of Physics, University of Sydney, NSW 2006, Australia\\
$^{2}$Australian Astronomical Optics, AAO-USydney, School of Physics, University of Sydney, NSW 2006, Australia\\
$^{3}$CAASTRO: ARC Centre of Excellence for All-sky Astrophysics\\
$^{4}$Department of Astronomy, University of Wisconsin-Madison, 475 North Charter Street, Madison, WI 53706, USA\\
$^{5}$ARC Centre of Excellence for All-sky Astrophysics in 3 Dimensions (ASTRO 3D)\\
$^{6}$School of Physics, University of New South Wales, NSW 2052, Australia\\
$^{7}$Australian Astronomical Optics, AAO-Macquarie, Faculty of Science and Engineering, Macquarie University, 105 Delhi Rd, North Ryde, NSW 2113, Australia\\
$^{8}$International Centre for Radio Astronomy Research, University of Western Australia, 35 Stirling Highway, Crawley, WA, 6009, Australia\\
$^{9}$Research School of Astronomy and Astrophysics, Australian National University, Canberra, ACT 2611, Australia\\
$^{10}$Atlassian 341 George St Sydney, NSW 2000\\
$^{11}$Department of Physics and Astronomy, Macquarie University, NSW 2109, Australia\\
$^{12}$Max Planck Institut f{\"u}r Astronomie, K{\"o}nigstuhl 17, 69117 Heidelberg, Germany\\
$^{13}$Ritter Astrophysical Research Center University of Toledo Toledo, OH 43606, USA\\
$^{14}$SOFIA Science Center, USRA, NASA Ames Research Center, Building N232, M/S 232-12, P.O. Box 1, Moffett Field, CA 94035-0001, USA\\
$^{15}$SUPA School of Physics \& Astronomy, University of St Andrews KY16 9SS Scotland\\
$^{16}$School of Physics University of Melbourne, Parkville, Victoria, Australia 3010\\
$^{17}$Department of Astronomy and Yonsei University Observatory, Yonsei University, Seoul 03722, Republic of Korea}
}

\pubyear{2018}

\begin{document}
\maketitle
\label{firstpage}

\begin{abstract}
We explore the radial distribution of star formation in galaxies in the SAMI Galaxy Survey as a function of their local group environment. Using a sample of galaxies in groups (with halo masses less than $ \simeq 10^{14} \, \mathrm{M_{\odot}}$) from the Galaxy And Mass Assembly Survey, we find signatures of environmental quenching in high-mass groups ($M_{G} > 10^{12.5} \,  \mathrm{M_{\odot}}$). The mean integrated specific star formation rate of star-forming galaxies in high-mass groups is lower than for galaxies in low-mass groups or that are ungrouped, with $\Delta \log(sSFR/\mathrm{yr^{-1}}) = ˆ'0.45 \pm 0.07$.  This difference is seen at all galaxy stellar masses. In high-mass groups, star-forming galaxies more massive than $M_{*} \sim 10^{10} \,  \mathrm{M_{\odot}}$ have centrally-concentrated star formation. These galaxies also lie below the star-formation main sequence, suggesting they may be undergoing outside-in quenching. Lower mass galaxies in high-mass groups do not show evidence of concentrated star formation. In groups less massive than $M_{G} = 10^{12.5} \, \mathrm{M_{\odot}}$ we do not observe these trends. In this regime we find a modest correlation between centrally-concentrated star formation and an enhancement in total star formation rate, consistent with triggered star formation in these galaxies.
\end{abstract}


\section{Introduction}
Obtaining a physical understanding of the modulation of star formation within galaxies remains an important goal for galaxy evolution studies.  Empirically, large galaxy surveys have uncovered broad trends that show star formation depends on both galaxy mass and environment \citep[e.g.][]{Peng2012}. Multiple measurements show that star formation is suppressed in high density environments \citep[e.g.][]{Lewis2002}, but the detailed physical nature of this suppression is not yet determined.

Processes such as ram pressure stripping or strangulation are thought to play a significant role.  Ram pressure stripping occurs when the kinetic interaction between the interstellar medium (ISM) of a galaxy in the intergalactic medium (IGM) forces the gas out of the galaxy \citep{Gunn72}.  Strangulation, sometimes referred to as starvation, occurs when the infall of gas into the disc of a galaxy is halted, starving the galaxy of fuel for future star formation \citep{Larson80}.  This process can occur when a galaxy falls into a cluster or group and its outer gaseous envelope is heated or removed.

Much of the literature concerning the impact of galaxy environments on star formation has focussed on studies of galaxy clusters. Narrow-band imaging studies of the distribution of H$\alpha$ emission in the Virgo cluster \citep[e.g.][]{Koopmann04a,Koopmann04b,Koopmann06} have shown that half of star-forming galaxies in this cluster have truncated star formation, with the star formation preferentially stopped in the outer parts of disks.  This is consistent with the observation of truncated neutral hydrogen gas disks in Virgo \citep{Cayatte90}. Truncation is also seen in dust \citep{Cortese2010} and ultraviolet star formation measurements \citep{Cortese2012}, while there are enhancements seen in central molecular gas \citep{Mok17}. The degree of truncation depends on the distance from the centre of Virgo \citep{Gavazzi2013}.

The above observations point to ram pressure being the dominant contributor to quenching in clusters such as Virgo, where it is particularly efficient for relatively low mass galaxies \citep[e.g.][]{Boselli2014,Boselli2016}.  
Comparing a suite of simulated galaxy orbits with observations of the projected phase-space distribution of galaxies from the SDSS, \citet{Oman2016} showed that in high--mass haloes ($>10^{13}$\,M$_{\odot}$) quenching is efficient.  However, others suggest lower efficiency \citep[e.g.][]{Wheeler2014}, albeit, for galaxies of lower mass.

Although clusters represent the most extreme environments, only $\sim5\%$ of galaxies exist in rich clusters. Because approximately $40-50 \%$ of galaxies exist in groups at $z \sim 0$ \citep{Eke04,Robotham2011}, the majority of environment-driven galaxy evolution is likely to occur outside of clusters \citep[e.g.][]{Balogh04}.  Even for cluster galaxies, it is likely that many are quenched prior to infall, via pre-processing in groups \citep[e.g.][]{Zabludoff1996,Fujita2004,Bianconi2018}.  Some of this pre-processing may be driven by ram pressure in filaments out to many times the virial radius of a cluster \citep{Bahe2013}.  This conclusion is consistent with HI stacking by \cite{Brown2017} that shows a deficit of neutral gas, even in relatively low mass groups ($10^{12} - 10^{13.5}$\,M$_\odot$).

Global analysis of star formation in large-scale spectroscopic galaxy surveys (e.g. the Sloan Digital Sky Survey; SDSS) points to environmental quenching being driven by processes that primarily act on satellite galaxies in haloes \citep[e.g.][]{Wetzel2012,Wetzel2013}. Satellite quenching may occur with a delay of a few Gyr after infall, followed by quenching on a timescale of  $<0.8$\,Gyr.  Others have reached the conclusion that the environmental quenching of star formation must be a rapid process \citep{Balogh04,Wijesinghe2011,Brough2013} based on the scarcity of galaxies in transition between star-forming and passive.  However, \cite{Rasmussen12} find that the integrated star formation rates of star-forming galaxies in groups are suppressed by $40$ per cent relative to galaxies outside of groups, so are likely in the process of being quenched. This result is in better agreement with work by \cite{vonderLinden2010} who study groups and clusters (typical halo mass of $\sim10^{14}$\,M$_{\odot}$) and also find star-forming galaxies that are transitioning.  The identification of a currently quenching population infers a relatively slow transition (several Gyr).  It is plausible that limits in sensitivity and/or aperture effects when using single fibre spectroscopy have made it difficult to identify galaxies currently in the process of quenching.  Taking an alternative approach, \cite{Peng15} use the difference in stellar metallicity between star-forming and passive galaxies to infer that slower starvation is preferred.

Spatially-resolved studies of environmental quenching outside of clusters are considerably more limited.  \cite{Cibinel2013} find that the outer parts of disc galaxies are relatively redder in high mass groups ($>10^{13.5}$\,M$_\odot$), compared to those in lower mass groups.  Qualitatively similar results are found by the narrow-band H$\alpha$ Galaxy Group Imaging Survey \citep[HAGGIS;][]{Kulkarni15}\footnote{Accessible at: \url{https://edoc.ub.uni-muenchen.de/18818/1/Kulkarni_Sandesh.pdf}}.  They find that galaxies below the star--formation main--sequence in groups typically have compact star formation with a steep radial profile.  In contrast, \cite{Eigenthaler2015} find no evidence of the truncation of star formation, as traced by H$\alpha$, in Hickson compact groups. These results suggest that radial truncation (presumably driven by ram pressure) is a factor in quenching over at least some range of halo masses in the group regime.

Other authors, using data from single-fibre spectroscopic surveys, have argued that much of the environment-driven evolution of galaxies can be explained by interactions between close pairs. \cite{Robotham14} showed that galaxies that are both dynamically and spatially close to their nearest neighbour are likely to have disturbed optical morphologies. This idea was expanded upon by \cite{Davies2015}, who showed that star formation in galaxies separated by less than $\sim 30 \, \mathrm{kpc}$ can also be affected. Their data showed that for galaxies in close pairs, the more massive galaxy tended to have its star formation enhanced, while the less massive galaxy had its star formation suppressed. They posited that the tidal disturbance of gas in the more massive galaxy would trigger star-formation. While the enhancement of star formation in close pairs was also reported in other studies \citep[e.g.][]{Ellison2008,Patton2013}, they did not study the suppression of star formation during these interactions. \cite{Davies2016a} showed that galaxies with stellar masses below $\sim 10^{8.5} \, \mathrm{M_{\odot}}$ become passive only in the presence of a more massive companion and argued that the increasing timescales for interaction between a galaxy of this mass and a more massive companion are consistent with their star formation being suppressed by strangulation. It is unclear from the research to date whether the environmental suppression of star formation in groups is due to galaxy-galaxy interactions or whether it can be attributed to the impact of the group environment at large.

Taken as a whole, the established literature suggests that in the most massive haloes ram pressure stripping is dominant, and that there is some evidence that ram pressure stripping continues to be important at halo masses in the group regime ($<10^{14}$\,M$_\odot$). A valuable route to diagnosing quenching mechanisms is spatially-resolved star formation measurements, and this is our focus for the current paper.  In previous work \citep{Schaefer17} we showed that as the local environment density (defined as 5th nearest neighbour) increases around a galaxy, the specific star formation rate (sSFR; $SFR/\mathrm{M_{*}}$) drops and this reduction in star formation occurs in the outer parts of the galaxy.  This infers that quenching occurs from the outside-in in the environments studied.  In the current paper we follow on from the results of \cite{Schaefer17}, studying how the spatial distribution of star formation changes in galaxies relative to physically motivated measures of local environment. In particular we focus on group properties, the location of galaxies within those groups and the estimated tidal force acting on each galaxy.  Our focus is on galaxies in haloes with mass less than $\sim10^{14}$\,M$_\odot$, as this is where our picture of environmental quenching is currently least clear.  

The remainder of this paper has the following layout. In Section \ref{Methods} we describe our data, discuss our sample selection and introduce the methods by which we measure the star formation properties of our galaxies. We compare these measurements of the star formation rate distribution to various metrics of local environment in Section \ref{Results}, discuss our findings in Section \ref{Discussion} and conclude with Section \ref{Conclusion}.

We assume a flat $\Lambda$CDM cosmology with $H_{0}=70$ km s$^{-1}$ Mpc$^{-1}$, $\Omega_{M}=0.27$ and $\Omega_{\Lambda}=0.73$. Unless otherwise stated, we adopt a \cite{Chabrier2003} stellar initial mass function for calculation of star formation rates.

\section{Methods}\label{Methods}
\subsection{SAMI Data}
The data for this study have been taken from the Sydney-Australian Astronomical Observatory Multi-object Integral Field Spectrograph \citep[SAMI;][]{Croom12} Galaxy Survey \citep{Bryant15} and the Galaxy and Mass Assembly \citep[GAMA;][]{Driver11,Hopkins13} survey. The SAMI Galaxy Survey is a resolved spectroscopic survey of over $3000$ galaxies performed using SAMI, which is mounted on the $3.9$\,m Anglo-Australian Telescope (AAT) at Siding Spring Observatory in Australia. SAMI comprises $13$ optical fibre hexabundles \citep{BlandHawthorn2011,Bryant2014} plugged into a steel plate at the prime focus of the AAT, $12$ of which are used to observe galaxies while the remaining hexabundle observes a standard star. These optical fibres feed into the AAOmega spectrograph, where the light is split into a red arm ($\lambda \lambda 6300-7400$\,\AA) and dispersed at a resolution of $R=4260$, and a blue arm ($\lambda \lambda 3700-5800$\,\AA) where it is dispersed to a resolution of $R=1810$ \citep[][]{vandeSande17}. The SAMI hexabundles are made of $61$ optical fibres fused to cover an approximately circular field of view with a $15\arcsec$ diameter on the sky. Within each hexabundle the optical fibres fill the aperture with an efficiency of $\sim73$\%. As a result, observations of galaxies with SAMI must be dithered to uniformly cover the image. We used approximately $7$ pointings of $1800$\,s integrations for a total $12600$\,s exposure. The raw data are reduced using the SAMI data reduction package, which has been written in the \textsc{python} language\footnote{Astrophysics Source Code Library, \\ ascl:1407.006 \url{ascl.net/1407.006}} and makes use of the 2dFDR pipeline \citep{Croom04}. The circular fibre cores are resampled onto a regular grid of $50 \times 50$ $0\farcs 5$ spaxels. For a full description of the data reduction, see \cite{Allen15} and for a discussion of representing the fibre data in a regularly gridded data cube see \cite{Sharp15}.

The galaxies observed for the main SAMI Galaxy Survey sample have been drawn from the equatorial regions of the GAMA spectroscopic survey (see Section \ref{GAMA_section}). The SAMI survey sample has a stepped selection function in stellar mass with redshift such that the final sample has a nearly uniform distribution of stellar masses. This sample selection covers a wide range of galaxy stellar masses ($10^{7} < M_{*}/M_{\odot} < 10^{11.5}$) in the redshift range $0.004 < z <0.11$ and includes galaxies in a wide variety of environments from non-group galaxies to galaxies in $10^{14} \, \mathrm{M_{\odot}}$ group halos. A thorough discussion of the SAMI target selection is given in \cite{Bryant15}. The SAMI survey augments the main GAMA-selected sample with a targeted sample of $\sim 800$ cluster galaxies \citep{Owers17}, chosen from the $2$ Degree Field Galaxy Redshift Survey \citep{Colless01} and the Sloan Digital Sky Survey \citep{York2000,Abazajian09}. At the time of this analysis, consistent measurements of the stellar masses and local environments surrounding the galaxies in the SAMI cluster sample were not available. This is due to the lack of highly complete spectroscopy and deep, multi-wavelength imaging as is available through the GAMA survey. For this reason the SAMI cluster galaxies have not been included for this work.
There is extensive existing literature on the quenching of star formation in galaxy clusters. However, it is not clear that the processes that act to quench galaxies in clusters will dominate in lower mass halos. For this reason, we focus on the environmental effects on galaxies in groups.  A study comparing the quenching mechanisms operating in clusters and groups will be presented in a future paper.

\subsection{GAMA Data} \label{GAMA_section}
GAMA, the parent survey for SAMI, is a deep, highly complete spectroscopic survey of galaxies made in three equatorial regions centred on $9$, $12$ and $15$ hours Right Ascension, with two additional non-equatorial fields that were not used for the SAMI selection. The equatorial fields have $98.5 \%$ complete spectroscopy to $r=19.8 \, \mathrm{mag}$, two magnitudes deeper than the SDSS \citep{Liske15}. 
\subsubsection{S{\'e}rsic photometry}
The GAMA survey targeted regions that have been covered by SDSS imaging in the $u$, $g$, $r$, $i$ and $z$ photometric bands. These images were re-analysed by \cite{Kelvin12}, who extracted objects from the images and fit single component S{\'e}rsic profiles to galaxies. We have made use of these data products, in particular the measurements of the effective radii ($R_{e}$), ellipticities and position angles extracted from the S{\'e}rsic fits to the SDSS $r$-band images.

\subsubsection{Stellar Masses}
We have used estimates of the stellar masses of the galaxies in our sample and their companions from version $18$ of the GAMA stellar mass catalogue. Stellar masses were computed by \cite{Taylor11} who used the \emph{ugriz} photometry and local-flow-corrected spectroscopic redshifts \citep{Tonry2000} to construct the rest-frame spectral energy distribution of each galaxy. These spectral energy distributions were used to model the stellar mass, star formation history, metallicity, and dust extinction in each galaxy assuming a \cite{Chabrier2003} stellar initial mass function (IMF). These stellar mass estimates are accurate to approximately $0.1$ dex for galaxies brighter than $r_{petro}=19.8$ mag.

\subsubsection{GAMA Galaxy Group Catalogue}
The deep and spectroscopically complete nature of the GAMA survey has allowed the creation of one of the most robust catalogues of galaxy groups made to date. \cite{Robotham2011} used a friends-of-friends linking algorithm to assign galaxies to groups. This two-step process uses both the projected separations of the galaxies and their redshifts to recover the true grouping of galaxies in space. The nature of the algorithm used is such that even pairs of galaxies are  identified as groups in the final catalogue. The grouping algorithm locates the central galaxy of a group and computes the group size, multiplicity (number of members above the detection limit), and velocity dispersion. From these measurements, it is possible to derive a number of properties of the group and its members including the total dynamical mass of the halo, the projected distance of each galaxy from the centre of the group, and the line-of-sight velocity of each galaxy with respect to the group centre. For an in-depth discussion of the group-finding algorithm used to derive the catalogue, see \cite{Robotham2011}, though note that at the time the original paper was published, the GAMA survey was still ongoing and consequently the size of the group catalogue and the spectroscopic completeness have since increased. We use the GAMA Galaxy Group Catalogue version $9$. The group halo masses and the associated uncertainties were calibrated by applying the group-finding algorithm to a set of mock catalogues from the Millennium Simulation \citep{Springel2005}. The uncertainty in group mass was found to be a function of the number of galaxies in each group ($N_{fof}$), $\log(M_{err}/M_{\odot})=1.0-0.43\log(N_{fof})$. This relation gives group mass errors of $\sim 0.87 \, \mathrm{dex}$ for the lowest mass groups, and $\sim 0.3 \, \mathrm{dex}$ for the most massive groups in our sample. A follow-up study of the GAMA groups catalogue using weak gravitational lensing demonstrated that the estimated group masses are unbiased down to at least $M_{G}=10^{13} \, \mathrm{M_{\odot}}$ \citep{Viola2015}. 

\begin{table*}
\caption{A description of the various samples used throughout this work, and the number of objects in each. The Inc-Pas and Final-Pas samples are the same since the identification of AGN-like emission becomes uncertain in galaxies with weak emission lines.}\label{sample_descriptions}
\begin{tabular}{c c p{8cm} c}
\hline
Name& Abbreviation & Description & Number\\ \hline 
Full Emission line & Full-Em & All galaxies with $\ewha>1 \, \mathrm{\AA}$ and $R_{e}<15 \arcsec$.& $1012$ \\
Full Passive & Full-Pas & All galaxies with $\ewha \leq 1 \, \mathrm{\AA}$ and $R_{e}<15 \arcsec$.& $246$ \\
Inc. Emission line & Inc-Em & All galaxies with $\ewha>1 \, \mathrm{\AA}$ and $R_{e}<15 \arcsec$, ellipticity $<0.7$, seeing$/R_{e} < 0.75$, seeing$<4\arcsec$, includes AGN-like central spectra & $477$ \\
Inc. passive & Inc-Pas & All galaxies with $\ewha \leq 1 \, \mathrm{\AA}$ and $R_{e}<15 \arcsec$, ellipticity $<0.7$, seeing$/R_{e} < 0.75$, seeing$<4\arcsec$ & $118$ \\
Final star-forming & Final-SF & All galaxies with $\ewha>1 \, \mathrm{\AA}$ and $R_{e}<15 \arcsec$, ellipticity $<0.7$, seeing$/R_{e} < 0.75$, seeing$<4\arcsec$, excluding AGN-like central spectra & $325$ \\
Final passive & Final-Pas & All galaxies with $\ewha \leq 1 \, \mathrm{\AA}$ and $R_{e}<15 \arcsec$, ellipticity $<0.7$, seeing$/R_{e} < 0.75$, seeing$<4\arcsec$ & $118$ \\
\hline
\end{tabular}
\end{table*}

\subsection{Sample Selection}\label{Sample_Selection}
We have selected galaxies from the SAMI Galaxy Survey in the GAMA regions. At the time of writing there were $1295$ galaxies from the SAMI main survey that have been observed and for which the data had been reduced. These galaxies comprise the SAMI internal data release v$0.9.1$. Our study of the spatially-resolved star formation properties of galaxies requires the selection of star-forming galaxies. Our method for selecting our star-forming sample mimics that of \cite{Schaefer17}. 

To measure whether a galaxy has any ongoing star formation, we have integrated the data cubes across both spatial axes and from the resulting spectrum we measure the equivalent width of the H$\alpha$ emission ($\ewha$), correcting for the underlying stellar absorption. If the absolute value of $\ewha$ is less than $1\, \mathrm{\AA}$, we say that the galaxy is passive. This H$\alpha$ equivalent width cutoff corresponds to a sSFR limit of approximately $10^{-12} \, \mathrm{yr^{-1}}$, and eliminates $253$ non-star-forming galaxies from our sample. In addition to this constraint we classify galaxies based on their emission line ratios in the central $2 \arcsec$. Based on the ratios of \mbox{[\ion{N}{ii}]} $\lambda 6583$ to H$\alpha$ and \mbox{[\ion{O}{iii}]} $\lambda 5007$ to H$\beta$ in this inner spectrum, we classify galaxies as either star-forming, composite or AGN/LINER based on their location on the \cite{BPT81} diagram. Galaxies with line ratios that place them above both the \cite{Kewley01} and \cite{Kauffmann03c} lines are classified as AGN/LINER, of which we find $333$ in our sample. Between these two constraints are a total of $165$ composite objects and below both lines are the star-forming objects.  Below we will refer to central spectra that are above the \cite{Kauffmann03c} line as AGN-like, noting that the emission is not necessarily due to accretion onto a super-massive black hole.

Recent advances have been made in decomposing the emission lines in galaxies hosting AGN into star-forming and AGN-excited components. \cite{DaviesAGN2016} proposed a technique whereby the lines are modeled as a linear combination of spectra from the uncontaminated AGN and HII regions within a given galaxy. Due to the redshift of the galaxies in our sample, this technique can not be applied. The physical resolution of SAMI is typically $\sim 1 \, \mathrm{kpc}$, meaning that a pure AGN basis spectrum cannot reliably be extracted from the data.

To reduce the effect of having hexabundles with a finite aperture on measuring the spatial distribution of star formation, we also limit the sizes of our galaxies that we include in our analysis. Twenty-six galaxies with effective radii greater than $15 \arcsec$ are rejected, as are $149$ galaxies for which the seeing of the observation is greater than $1 \, R_{e}$. We also excluded galaxies that have ellipticity values greater than $0.7$ to eliminate $200$ edge-on systems. Note that some galaxies have been rejected based on more than one constraint.

Our final star-forming sample comprises \SFNUM\ galaxies, including \ISONUM\ galaxies not assigned to groups and \GNUM\ galaxies in groups.  For here onwards we will refer to this sample as the final star-forming sample (or Final-SF).  We will refer to the sample of galaxies that meet all other cuts, but that have $\ewha<1$\AA\ as our passive sample (Final-Pas). We show the distribution of halo masses for our final star-forming sample in Fig.\ \ref{G_mass_hist} that covers a range of masses between $10^{9}$ and $10^{14.5} \, \mathrm{M_{\odot}}$. These histograms do not include galaxies that have not been identified as belonging to a group in the GAMA Galaxy Group Catalogue.

Throughout this paper we utilise several different subsamples, using the selections described above. These allow us to test the robustness of our conclusions to changes in our selection criteria.  These are summarized in Table \ref{sample_descriptions}.  The full samples (Full-Em and Full-Pas for those above and below the $\ewha=1$\,\AA\ limit respectively) contain almost all galaxies, only removing a very small fraction with $R_e>15 \arcsec$.  The Inc-Em and Inc-Pas samples also have the inclination and size cuts applied (but not any rejection of central AGN-like emission).  The Final-SF and Final-Pas samples also apply the constraint removing objects with central AGN-like spectra.

\subsubsection{Possible effects of excluding AGN-like galaxies}\label{logit_regression}
The sample selection criteria introduced in Section \ref{Sample_Selection} were devised to provide a sample of galaxies where our measurements of the spatial distribution of star formation are robust.  As a result, the Final-SF sample is a relatively small fraction of the entire sample. Some galaxies that have been removed from our sample failed to satisfy multiple criteria for selection. In particular, of the $333$ galaxies that were identified as having AGN-like emission in their centres, $203$ of them simultaneously failed our integrated $\ewha$ cut. Given the number of galaxies not included, it is important to understand whether our cuts have biased the final star-forming sample.

\begin{figure}
\includegraphics[scale=0.9]{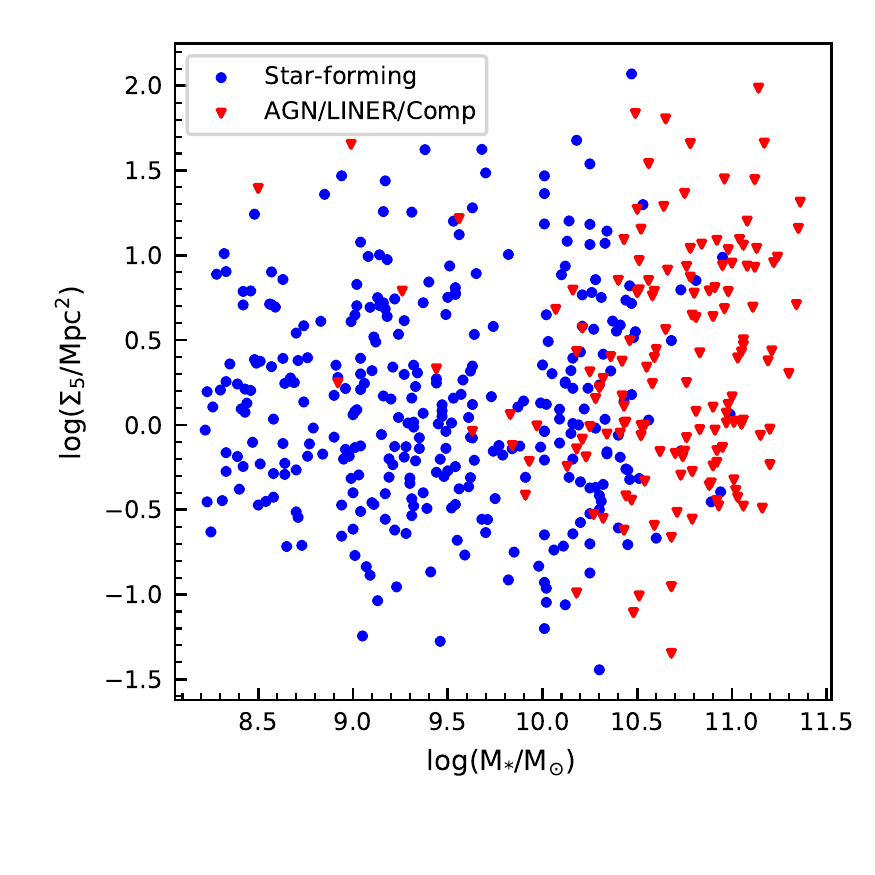}
\caption{The distribution of stellar mass vs.\ 5th-nearest-neighbour density for galaxies that are selected in our star forming sample (blue circles) and those that have been rejected due to central AGN-like emission (red triangles). The separation between star formation and AGN-like emission is only dependent on stellar mass.}\label{mass_sig5}
\end{figure}

\begin{table*}
\begin{center}
\begin{tabular}{rcccc}
\hline
&Coeff. & Standard error & $z$-value  & $P(z = 0)$\\
&(1)&(2)&(3)&(4) \\\hline
Intercept &$-36.5$&$3.6$&$-10.1$&$4.3\times10^{-24}$\\
$\log(M_{*}/\mathrm{M_{\odot}})$&$3.50$&$0.35$&$10.1$&$8.0\times 10^{-24}$\\
$\log(\Sigma_{5}/\mathrm{Mpc^{2}})$&$0.25$&$0.22$&$1.16$&$0.25$\\
\hline
\end{tabular}
\end{center}
\caption{The results of a logistic regression to determine the probability of a galaxy having a central AGN-like spectrum as a function of stellar mass and environment density. Column (1) shows the regression coefficients that represent the change in the log odds of a galaxy hosting a central AGN-like spectrum. Column (2) is the standard error on the coefficients, column (3) is the ratio of the standard error to the coefficient, and column (4) is the p-value against the null hypothesis. The probability of a galaxy hosting central AGN-like emission is correlated with stellar mass, but not significantly with environment.}\label{logit_table}
\end{table*}

Galaxies with $\ewha>1$\,\AA\ tend to have high stellar masses and preferentially occupy dense environments. As was noted above, of the $253$ galaxies that did not satisfy our $\ewha$ criterion, $203$ have central emission line ratios that classify them as AGN-like.  This is largely because quenched galaxies often have weak LINER-like emission. Nevertheless, it is possible that we are preferentially excluding galaxies in dense environments based on their central emission line ratios that would otherwise have detectable star-formation. We test for this possibility using a logistic regression \citep{Cox1958,Walker1967} that models the probability of a galaxy having central AGN-like emission line ratios as a function of $\log(M_{*}/\mathrm{M_{\odot}})$ and fifth-nearest neighbour environment surface density $\log(\mathrm{\Sigma_{5}/Mpc^{2}})$ as measured by the GAMA survey \citep{Brough2013}. We use $\mathrm{\Sigma_{5}}$ to quantify the environment in this context because it provides a measure of the environment density of galaxies that can be measured for our entire sample, including galaxies not found in groups. Within our group sample, $\Sigma_{5}$ is correlated with the group mass, and other metrics of environment density. While the following test utilizes $\mathrm{\Sigma_{5}}$ as a probe of environment, the result is therefore applicable to other measures of environment (in fact the same result is found if we use group mass rather than $\mathrm{\Sigma_{5}}$ for only the group galaxies).

Our test uses the $477$ galaxies in the Inc-Em sample (see Table \ref{sample_descriptions}.  These have that have integrated $\ewha$ larger than $1 \, \mathrm{\AA}$ and satisfy all of the data quality constraints imposed on our sample (PSF FWHM$/R_{r}$ $<0.75$, ellipticity $< 0.7$, $R_{e}<15\arcsec$, and the environment density flag in the GAMA catalogue is set to $0$) and include galaxies with AGN-like central spectra.  The distribution of these galaxies can be seen in Fig.\ \ref{mass_sig5}.  A clear division between the objects with central AGN-like spectra (red points) and those that are star-forming (blue points) can be seen.  This division is a strong function of stellar mass, but there is no visible separation by environment.  The results of the logistic regression confirm this and can be found in Table \ref{logit_table}.  This analysis finds that within our sample the probability of a galaxy hosting AGN-like central emission line ratios is strongly correlated with its stellar mass (correlation coefficient $3.50\pm0.35$), but that there is no significant relationship with $\Sigma_{5}$ (correlation coefficient $0.25\pm0.22$)\footnote{These coefficients can be interpreted as the increase in the log-odds, $\mathrm{ln}(p/(1-p))$, of a galaxy hosting central AGN/LINER/composite emission line ratios per unit increase in $\mathrm{log}({M_{*}/\mathrm{M_{\odot}}})$ or $\mathrm{log(\Sigma_{5}/Mpc^{2})}$.}. Thus, our removal of AGN (in order to measure robust star formation morphologies) does not introduce a bias that depends on environment.  It does introduce a bias as a function of stellar mass. To negate this we will, where possible, compare results as a function of mass.  While largely focusing on the Final-SF sample only for the remainder of this paper, we will discuss the impact of leaving central AGN-like galaxies in the sample at various points.

\begin{figure}
\includegraphics[scale=0.9]{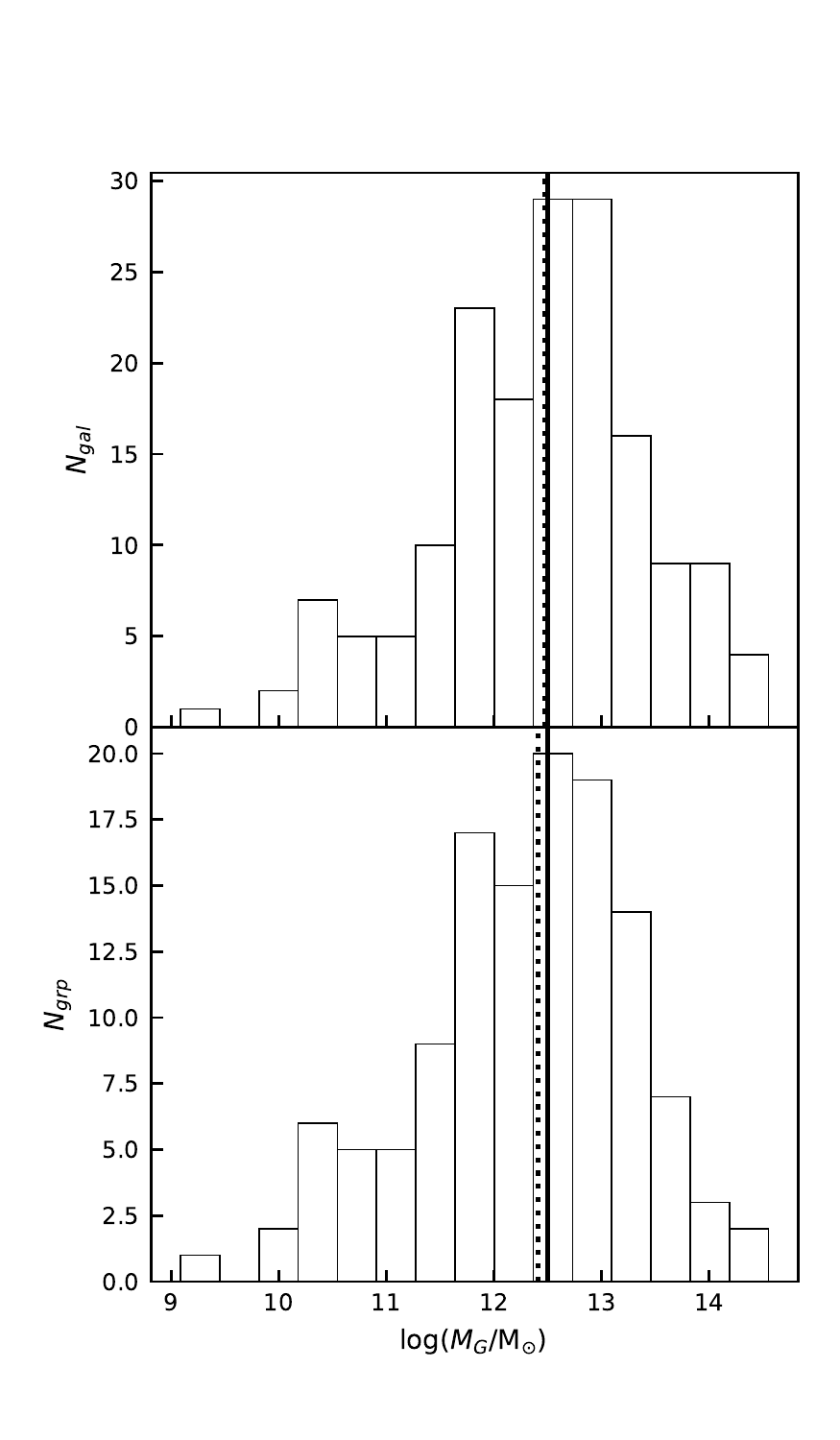}
\caption{The distribution of halo masses for galaxies detected as being members of groups in our final star-forming sample. In the upper histogram we show the number of galaxies in each interval of group mass, while in the lower histogram we show the number of individual groups in each interval of group mass. The vertical black solid line marks the $10^{12.5} \, \mathrm{M_{\odot}}$ division used throughout this work. The dotted line shows the median group mass, $M_{G} = 10^{12.46} \, \mathrm{M_{\odot}}$.}\label{G_mass_hist}
\end{figure}

\subsubsection{The distribution of galaxy apparent sizes as a function of stellar mass}
The selection of our Final-SF sample included the criterion that the PSF FWHM for an observation must not exceed $0.75$ times the $r$-band effective radius of the galaxy. This constraint was imposed to ensure that the measurement of the spatial extent of star formation is as robust to PSF effects as possible. However, the possibility exists that this criterion may adversely affect our coverage of low-mass galaxies. Splitting the full sample into three evenly-spaced bins of stellar mass ($\log(M_{*}/\mathrm{M_{\odot}})<9.1$, $9.1\leq \log(M_{*}/\mathrm{M_{\odot}}) < 10.1$, and $\log(M_{*}/\mathrm{M_{\odot}}) \geq 10.1$), we compare the means of the $r$-band effective radii in these ranges. These values are $4.65\pm0.38$, $5.03\pm0.35$, and $4.16\pm0.47$ arcsec for the low, intermediate and high stellar mass bins respectively. These are broadly consistent with each other thanks to the selection of the main SAMI Galaxy Survey that has a stepped selection function in stellar mass and redshift. The conclusion that we draw from this is that the removal of galaxies that are small relative to the PSF will not strongly affect low mass galaxies any more than high-mass galaxies, and consequently any trends inferred to be a function of stellar mass are not impacted by this selection criterion.

\subsubsection{Comparison between different group environments}

Interpreting any variation in the star-forming properties of galaxies between different subsamples requires that there be no major discrepancies in the galaxy properties between the different subsamples. In Section \ref{Results} we introduce three subsamples: i) galaxies that are not in groups, ii) galaxies that are in the \cite{Robotham2011} group catalogue with halo masses below $10^{12.5} \, \mathrm{M_{\odot}}$, and iii) galaxies in groups with halo masses greater than $10^{12.5} \, \mathrm{M_{\odot}}$.  Fig. \ref{G_mass_hist} shows the distribution of group halo masses.  The mean halo masses for the samples above and below our $10^{12.5} \, \mathrm{M_{\odot}}$ threshold are $10^{13.2} \, \mathrm{M_{\odot}}$ and $10^{11.7} \, \mathrm{M_{\odot}}$ respectively.  It should be noted that there are only a handful of galaxies included that exist in haloes of mass greater than $10^{14} \, \mathrm{M_{\odot}}$.  Therefore our results reflect the role of group rather than cluster environments (where the nominal dividing line between the two is typically taken to be $10^{14} \, \mathrm{M_{\odot}}$).

For the star--forming galaxies that satisfy the selection criteria outlined above, a K-S test indicates no statistically significant differences in the distributions of stellar mass, effective radius or redshift between the three environment samples ($D_{\rm KS}< 0.2$ and $p>0.3$ between each variable in each group mass bin).  However, given that errors in group mass can be large for low-mass haloes we expect some overlap in halo mass between the ungrouped and low--mass group sample.  We compare the environments in these two samples using the nth--nearest neighbour density [with the velocity limit of $1000$\,km\,s$^{-1}$ and the same density defining sample as used by \citet{Schaefer17}]. These surface density measurements are described in full by \cite{Brough2013}. Using a K-S test, the nth--nearest neighbour density is found to be significantly different between the ungrouped and low--mass group sample for all $n$ values tested: $n=5$ ($\Sigma_5$, $D_{\rm KS}=0.20$, $p=0.02$), $n=3$ ($\Sigma_{3}$; $D_{\rm KS}=0.26$, $p=0.0005$) and $n=1$ ($\Sigma_{1}$; $D_{\rm KS}=0.533$, $p=10^{-14}$). The increased significance for smaller $n$ suggests that the low--mass group is dominated by groups of low multiplicity, including pairs.

Given that the mass, redshift and size distribution of our galaxies is the same for all samples, and that the GAMA spectroscopic completeness is over $98$ per cent, it is clear that although there will be some overlap in halo mass between the ungrouped and low--mass group sample, on average they correspond to different environments.

\subsection{Analysis of SAMI data}
The analysis of the SAMI data is as described in \cite{Schaefer17}, but for completeness we summarise the process here. 
\subsubsection{Annular Voronoi Binning}
To facilitate a robust correction for dust attenuation along the line of sight, we applied annular Voronoi binning to the SAMI data cubes. Spaxels are added together in elliptical annuli to a signal-to-noise ratio of $10$ per \AA ~in the continuum at the wavelength of the H$\beta$ line. An adaptive binning scheme has the advantage of providing sufficient signal to allow the subtraction of the H$\beta$ absorption line and thus an accurate correction for dust extinction. This binning additionally ensures that the spatial scale over which a single dust correction is applied is minimised, while further ensuring that the radial structure in each galaxy is preserved. We have binned the SAMI data in $0\farcs 5$-wide elliptical annuli that are defined by the ellipticity and position angle obtained from the GAMA S{\'e}rsic photometry. This technique is outlined in \cite{Schaefer17}.

\subsubsection{Spectral fitting with \textsc{lzifu}}
We fitted the spectrum within each annular Voronoi bin using \textsc{lzifu} \citep{Ho2016}. \textsc{lzifu} is a spectral fitting pipeline written in the Interactive Data Language (IDL). It utilises the Penalised Pixel Fitting algorithm \citep[pPXF;][]{Cappellari04} to model the stellar continuum light from each galaxy. For each spectrum we fitted a linear combination of simple stellar population (SSP) models from the MILES library \citep{Vazdekis10} with an 8th degree multiplicative polynomial to take into account any residual flux calibration errors and the reddening of the continuum from astrophysical sources. We used a $65$ template subset of the full MILES library. This subset covers five metallicities from $ [\mathrm{Z/H}] = -1.71$ to $0.22$ and thirteen ages spaced logarithmically in the range $0.063-14$ Gyr. The continuum model derived by pPXF was subtracted from the data, and the emission lines, including H$\alpha$ and H$\beta$, were fitted with single component Gaussians using the \textsc{mpfit} routine \citep{Markwardt09}.

\subsection{Star-forming properties of galaxies}
We use a number of metrics to determine the impact of the group environment on the star formation in the galaxies in our sample.

\subsubsection{Integrated star formation rates}
We calculated the total star formation rate within the SAMI aperture by adding the dust-corrected flux from each annular Voronoi bin. Dust extinction corrections are applied by measuring the departure of the Balmer decrement, the ratio of measured H$\alpha$ flux to H$\beta$ flux ($BD$; 
$f_{\mathrm{H}\alpha}/f_{\mathrm{H}\beta}$), from the canonical value of $2.86$ predicted for Case B recombination under standard conditions.
Under the assumption that the intervening dust forms a foreground screen to the HII regions in our target galaxies \citep{Calzetti01} and using the \cite{Cardelli89} dust extinction curve, the obscuration-corrected H$\alpha$ flux is
\begin{equation}
F_{\mathrm{H}\alpha}=f_{\mathrm{H\alpha}}\left(\frac{BD}{2.86}\right)^{2.36}
\end{equation}
in each spectrum. In cases where the signal-to-noise ratio for the H$\beta$ emission line is less than $3$, or the measured Balmer decrement is less than $2.86$, we assume no dust extinction and use the raw H$\alpha$ flux. Spaxels with emission lines not produced by star formation, as determined by their location on the BPT diagram, are rejected. Integrating the dust-corrected fluxes over the SAMI aperture gives the integrated flux, which is converted to a luminosity using the redshift of each galaxy
\begin{equation}
L(\mathrm{H}\alpha)=\frac{F_{\mathrm{H}\alpha}}{4 \pi d_{L}^{2}},
\end{equation}
where $d_{L}$ is the luminosity distance to the galaxy. We calculate the star formation rates in our galaxies with the \cite{Kennicutt98} relation assuming a \cite{Chabrier2003} IMF:
\begin{equation}\label{SFR_eqn}
\mathrm{SFR}=\frac{L_{\mathrm{H}\alpha}~(W)}{2.16\times 10^{34}} ~\mathrm{M}_{\odot} ~\mathrm{yr}^{-1}.
\end{equation}

The estimated uncertainty on these star-formation rates consists of several components. The random error from fluxes in each annular Voronoi bin are combined and contribute an average uncertainty of $0.015 \, \mathrm{dex}$ to the star formation rate measurements. The dominant component of our error budget comes from uncertainties in the calibration of Equation \ref{SFR_eqn}. \cite{Kewley2002} measured a scatter of $10$ per cent between the H$\alpha$ and infrared star formation rates for an error contribution of $0.05 \, \mathrm{dex}$. Combining these uncertainties with those of the stellar masses gives specific star formation rate errors that average $0.12 \, \mathrm{dex}$.

\subsubsection{The spatial distribution of star formation}\label{sec:r50}

We quantify the radial extent of star formation in galaxies within our sample by making use of the ratio $r_{50,\mathrm{H\alpha}}/r_{50,cont}$, described in \cite{Schaefer17}. This measurement compares the radius within which half of the dust-corrected H$\alpha$ emission emanates to the radius containing half of the continuum light from the part of the galaxy that lies within the view of the SAMI hexabundles. These radii are calculated by measuring the curve-of-growth for the emission or continuum light. In calculating the curves-of-growth we have made the assumption that galaxies in our sample are idealised thin discs and any ellipticity is due to their inclination to our line of sight. The ellipticity and position angle on the sky are taken from the GAMA S{\`e}rsic photometric fits to SDSS $r$-band images. An in-depth discussion of the measurement, advantages of, and systematic effects that can arise by making this measurement on galaxies observed with $15 \arcsec$ integral field units can be found in \cite{Schaefer17}. 

We estimate the uncertainty on this quantity using a Monte Carlo approach. The measured values of the H$\alpha$ and H$\beta$ fluxes, and the estimated galaxy ellipticities and position angles, were shifted by a random amount corresponding to the measurement errors on each quantity. The error distributions were assumed to be Gaussian. This process was repeated $1000$ times for each galaxy, and $r_{50,\mathrm{H\alpha}}/r_{50,cont}$ was remeasured. The error on this quantity is calculated as the standard deviation of the resulting distribution of measured values. For the galaxies in our final star-forming sample, this typically results in an error of $0.02$ on $r_{50,\mathrm{H\alpha}}/r_{50,cont}$, the majority of which is accounted for by the errors on the ellipticity and position angle of the galaxy. 

Another source of potential uncertainty is the effect of the seeing on our measurement of the spatial extent of star formation. The impact of seeing is to convolve the H$\alpha$ and continuum images of the galaxy with the point spread function. If the ratio $r_{50,\mathrm{H\alpha}}/r_{50,cont}$ is intrinsically close to $1$, then convolution with the PSF will not lead to a large change in the observed ratio. This is because the numerator and denominator of the fraction will change by roughly the same amount. If H$\alpha$ is more extended than the continuum, the fractional increase in $r_{50,\mathrm{H\alpha}}$ will be smaller than for $r_{50,cont}$ after the convolution. If H$\alpha$ is less extended than the continuum, then the fractional increase in $r_{50,\mathrm{H\alpha}}$ will be larger than for $r_{50,cont}$. Thus, the effect of the seeing is to force the observed value of $r_{50,\mathrm{H\alpha}}/r_{50,cont}$ towards $1$.

Using a suite of toy model galaxies, in which we simulated the H$\alpha$ and continuum light as simple Gaussian distributions with varying sizes, we investigated the magnitude of this bias. The images of these model galaxies were convolved with a $2\farcs 5$ Gaussian PSF. When the input $\log(r_{50,\mathrm{H\alpha}}/r_{50,cont})$ was between $-0.2$ and $0.2$, the seeing changed the measured value by $0.04$ on average after convolution. For galaxies with very centrally concentrated star-formation, this systematic can be larger, depending on the intrinsic size of the $H\alpha$ distribution. The effect of seeing is therefore to reduce the scatter in the measured value of $r_{50,\mathrm{H\alpha}}/r_{50,cont}$. While the slope of any trends may be affected slightly, we will make use of Spearman rank correlation tests where appropriate, as this statistic is less sensitive to the change in slope of a correlation in this situation.

As explained in \cite{Schaefer17}, the largest source of uncertainty is the amount of star formation outside of the SAMI aperture. There is no systematic difference in the radial coverage of galaxies between environments in our sample.

\begin{figure*}
\includegraphics{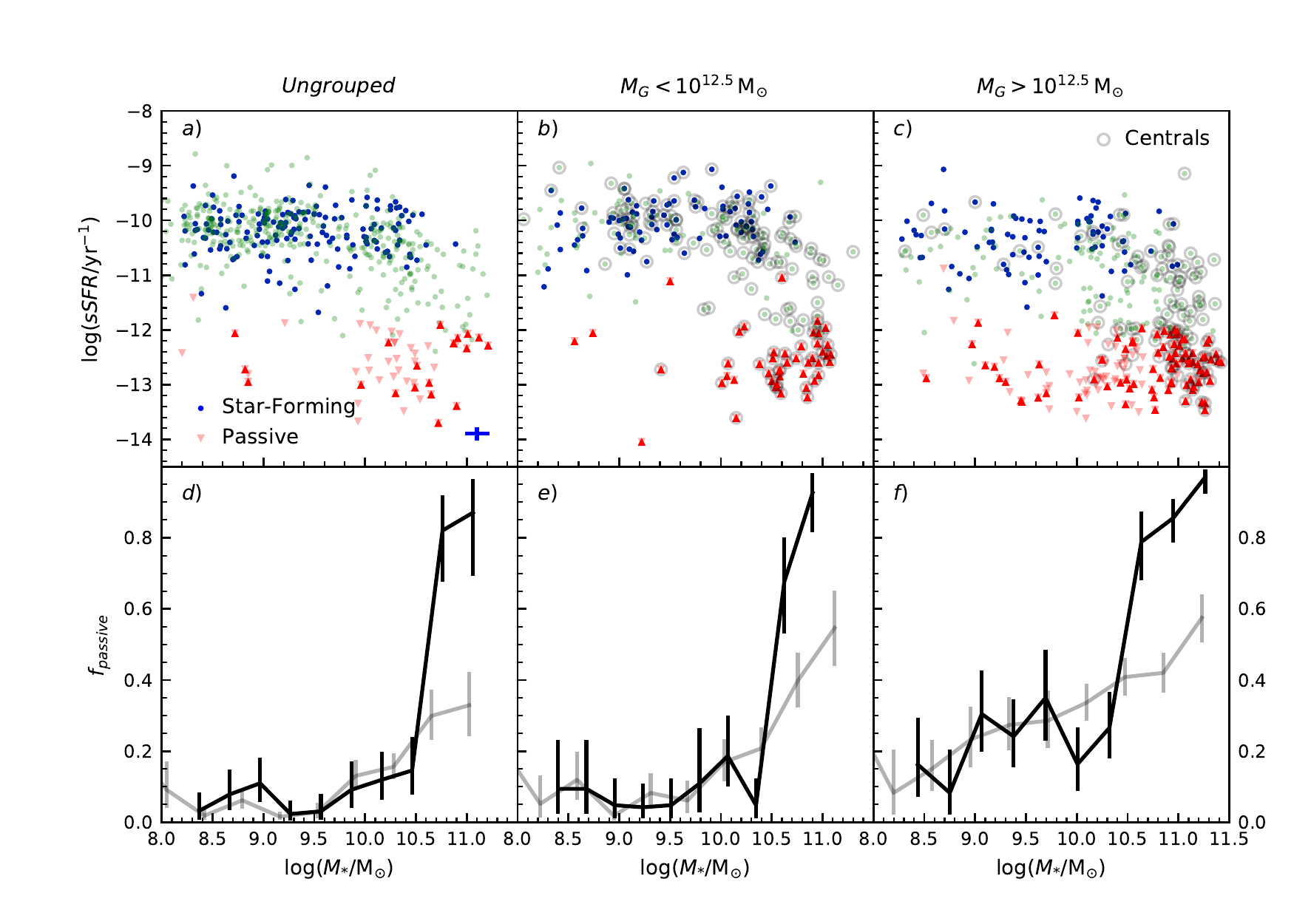}
\caption{The top row (panels $a-c$) shows the sSFR of star-forming galaxies in our final star-forming sample (blue points) and final passive sample (red triangles) as a function of stellar mass, $\log(M_{*}/M_{\odot})$, in ungrouped galaxies (left), groups with halo mass less than $10^{12.5} \, \mathrm{M}_{\odot}$ (centre), and groups more massive than $10^{12.5} \, \mathrm{M_{\odot}}$ (right). Markers surrounded by grey circles represent galaxies that are the centrals of their halo. In addition to galaxies in the Final-SF sample, we also show the specific star formation rate estimates for the Full-SF and Full-Pas galaxies in light green and light red respectively. The typical error on the stellar mass and star formation rates for star-forming galaxies are shown at the lower right of panel $a$. The lower row (panels $d-f$) shows the fraction of galaxies that were classified as passive in bins of stellar mass for galaxies satisfying our final selection criteria (black) and the full SAMI sample (grey).}\label{logmstar_q_frac_G}
\end{figure*}

\section{Results}\label{Results}
\subsection{Group mass and integrated star formation rates}
There is a significant body of work in the literature that focuses on how the star formation rates of galaxies and the fraction of passive galaxies change with stellar mass and environment density \citep[e.g.][]{Peng2010,Wijesinghe2011,Wijesinghe12,Peng2012,Wetzel2012,Alpaslan2015,Davies2016a}. These results are generally based on large-scale single-fibre spectroscopic or photometric surveys. With a representative sample of \SFNUM\ star-forming galaxies, our primary aim in this paper is not to duplicate previous results from single fibre surveys, but to use the spatially-resolved star formation measurements from SAMI to investigate how the location of star formation is modulated by environment.  However, in this section we will briefly discuss our integrated star formation rates and examine whether there are any trends with environment.

\begin{figure*}
\includegraphics[width = 18cm]{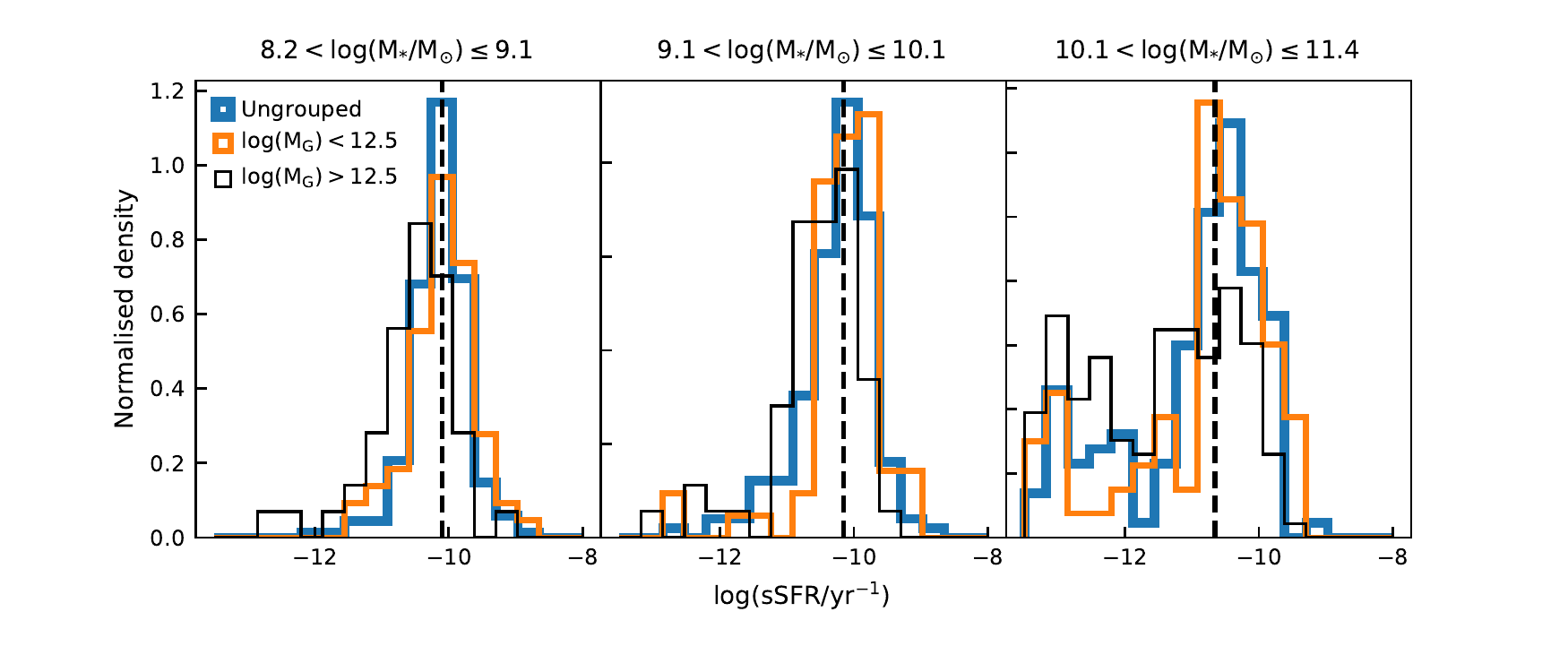}
\caption{The normalised distribution of sSFRs for ungrouped galaxies (blue), galaxies in groups with $\mathrm{M}_{G} < 10^{12.5} \, \mathrm{M_{\odot}}$ (orange) and galaxies in groups with $\mathrm{M}_{G} > 10^{12.5} \, \mathrm{M_{\odot}}$ (black).  We show this separately for three different intervals of galaxy stellar mass.  This includes galaxies that have central AGN-like emission (using the Full-Em sample), but only star-forming spaxels are included in the estimation of sSFR.  The vertical, black dashed line represents the median sSFR for the ungrouped galaxies in each stellar mass interval. In all three stellar mass intervals galaxies in the higher mass groups have systematically lower sSFR.}\label{SF_MS_Gmass_hist}
\end{figure*}

We quantify the effect of group environments on the star-forming properties of galaxies by comparing the sSFRs of galaxies to the masses of their parent group haloes. In Fig.\ \ref{logmstar_q_frac_G} we split our entire sample into three intervals of halo mass. A non-grouped sample, which comprises galaxies that do not appear in the GAMA Galaxy Group Catalogue \citep{Robotham2011}, a low-mass group sample with groups that have multiple galaxies within halos of mass below $10^{12.5} \, \mathrm{M_{\odot}}$, and a high-mass group sample with galaxies in group halos more massive than $10^{12.5} \, \mathrm{M_{\odot}}$. This boundary was chosen to approximately evenly split the grouped galaxy sample in two, and corresponds approximately to the threshold mass at which environmental effects are seen in other surveys \citep[e.g.][]{Rudnick2017}. The mean halo mass in each subsample is $10^{13.2} \, \mathrm{M_{\odot}}$ for galaxies in the high-mass groups, and $10^{11.7} \, \mathrm{M_{\odot}}$ for galaxies in the low-mass groups. We note that group masses here are derived from the dynamics of galaxies within each halo. For galaxies in the ungrouped subsample, the method of \cite{Robotham2011} is unable to estimate the halo mass. Based on the tight correlation between total stellar mass within a group and its halo mass presented by \cite{Yang2007}, we estimate that the most massive halos in the ungrouped sample will be of order $10^{11} - 10^{12} \, \mathrm{M_{\odot}}$, but are unlikely to host many satellite galaxies brighter than the GAMA detection limit. We will not use this kind of estimate for the remainder of this paper. 

In the upper row of panels of Figure \ref{logmstar_q_frac_G}, we show the sSFRs of our star-forming sample using blue circles, and of our passive galaxy sample with red triangles. Objects that are not in the final samples (Final-SF or Final-Pas) are shown as green (for $\ewha$ $>1$\AA) and light red (for $\ewha <1 \, \mathrm{\AA}$), these include galaxies that have been removed due to central AGN-like emission, or due to size/inclination cuts. At low stellar mass (below $\sim10^{10} \,\mathrm{M_{\odot}}$) the distribution of sSFR is the same for both our star-forming sample and the galaxies rejected.  However, at higher stellar mass, galaxies that have been rejected are preferentially at lower sSFR.  This is because of the increasing contribution from objects with central AGN-like emission as stellar mass increases.

In all environments, there is an increase in the passive fraction of galaxies that satisfy our selection criteria at a stellar mass of approximately $10^{10.5} \, \mathrm{M_{\odot}}$, in agreement with numerous previous studies \citep[e.g.][]{Kauffmann2003b,Geha2012}. The passive fraction based on galaxies that satisfy our final selection criteria is shown by the solid black lines in the lower panels of Figure \ref{logmstar_q_frac_G}. We note that if all galaxies in the SAMI sample are included, the rise in passive fraction with stellar mass is more gradual (grey lines). This more gradual trend is consistent with the results presented by \cite{Wetzel2012}, and is driven by objects that are rejected on the basis of having central AGN-like emission. Overall, our passive fraction for the entire sample is somewhat lower than \cite{Wetzel2012}. The cause of this discrepancy is likely to be because \cite{Wetzel2012} define a galaxy as passive if its specific star formation rate is below $10^{-11} \, \mathrm{yr}^{-1}$, while our integrated H$\alpha$ equivalent width limit corresponds to a specific star formation rate of approximately a factor of ten lower.

When using the full sample (Full-Em and Full-Pas) we find a difference in the passive fraction of low mass galaxies (below $10^{10.5} \, \mathrm{M_{\odot}}$).  The passive fraction is $7 \pm 1$ percent, $7\pm 1$ percent and $29 \pm 3$ percent for the ungrouped, low group mass and high group mass samples respectively.  While calculating the passive fraction of galaxies is not the main focus of this work, our results are qualitatively consistent with results found in the literature.

In addition to the increased passive fraction of galaxies in the most massive groups, the sSFRs of star-forming galaxies are reduced in dense environments. This is shown by the sSFR distributions in Fig.\ref{SF_MS_Gmass_hist}, measured in 3 independent stellar mass intervals (chosen to have approximately equal number of galaxies from our final star-forming sample).
Here we use all galaxies with $\ewha>1$\,\AA\ (i.e. the Full-EM sample, which includes all blue and green points in Fig.\ \ref{logmstar_q_frac_G}) as we are just concerned with integrated values, not whether we can resolve structure. The mean sSFRs for each environment and stellar mass interval are listed in Table \ref{tab:ssfr}.  In every stellar mass interval we see a significant decrease in the mean sSFR in the highest mass groups, with the difference between the ungrouped and high-mass groups being $\Delta \log(sSFR/{\rm yr}^{-1}) = -0.38\pm0.10$, $-0.41\pm0.13$ and $-0.64\pm0.14$ for the stellar mass intervals $M_*<10^{9.1}$\,M$_\odot$, $10^{9.1}<M_*<10^{10.1}$\,M$_\odot$ and $M_*>10^{10.1}$\,M$_\odot$ respectively.  If we use all spaxels (including AGN--like spaxels) we find the same differential trends (although sSFR can be biased high by the non-SF contribution, we find the same bias in different environments, leading to the same differential between environments).  Likewise, we find the same result if we restrict our sample to just our Final-SF sample (blue points in Fig. \ref{logmstar_q_frac_G}), although the smaller number of galaxies results in larger uncertainties.

\begin{table*}
\begin{center}
\caption{The mean sSFR for all galaxies with $\ewha>1$\,\AA\ as a function of stellar mass and environment.  This includes galaxies that have central AGN/LINER/composite emission, but only star-forming spaxels are included in the estimation of sSFR.  sSFRs are given for 3 intervals of stellar mass and group mass ($M_G$), as well as ungrouped galaxies.}\label{tab:ssfr}
\begin{tabular}{lccc}
\hline
Stellar mass &  \multicolumn{3}{c}{$\log(sSFR/{\rm yr}^{-1})$}  \\
 & Ungrouped & $M_{G} < 10^{12.5}$& $M_{G} > 10^{12.5}$\\
\hline 
$\log(M_*/M_\odot)<9.1$ & $-10.10\pm0.03$ & $-10.04\pm0.06$ & $-10.49\pm0.10$ \\
$9.1<\log(M_*/M_\odot)<10.1$ & $-10.28\pm0.05$ & $-10.20\pm0.09$ & $-10.69\pm0.12$ \\
$\log(M_*/M_\odot)>10.1$ & $-11.10\pm0.09$ & $-10.99\pm0.12$ & $-11.66\pm0.10$ \\
\hline
\end{tabular}
\end{center}
\end{table*}

Fig.\ \ref{SF_MS_Gmass_hist} shows that the sSFR distribution changes in two ways as group mass increases.  First, there is a shift of the peak sSFR.  This is particularly noticeable at low/intermediate stellar mass.  Second, there is an increasing number of galaxies in the low sSFR tail as we move to higher mass groups.  The exact number of galaxies in the low sSFR tail will depend on the threshold used to define the star-forming sample and this will influence the measured mean sSFR.  However, given that we are using the full galaxy sample and that we have shown earlier that there are no environmentally-dependent biases in the sample selection, this shows that star-forming galaxies in a high-mass group environment will have lower sSFRs.  At low/intermediate stellar mass ($\log(M_*/M_\odot)<10.1$) the contribution of AGN is also small (see Fig.\ \ref{mass_sig5}), adding further to the confidence of this conclusion.

\begin{figure*}
\includegraphics{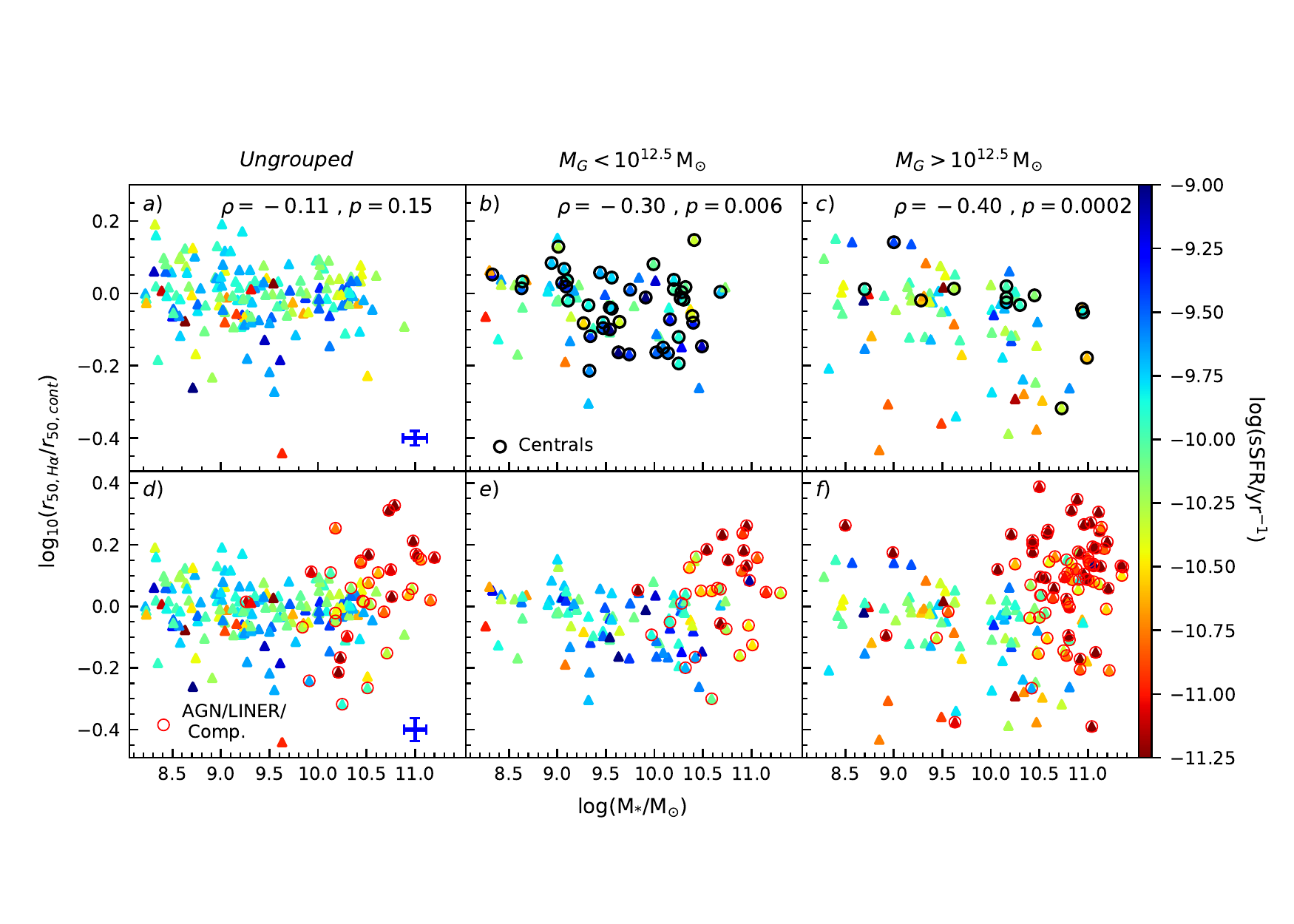}
\caption{The scale-radius ratio as a function of stellar mass in different intervals of galaxy group halo mass, with each point coloured by the specific star formation rate. The top row (panels $a-c$) shows only galaxies in our Final-SF sample. In the upper right of each panel we show the Spearman rank correlation coefficient, $\rho$, and the associated p-value. Representative errors on each quantity are shown in blue in the lower right of panel $a)$. For the ungrouped sample there is no statistically significant correlation, but in progressively more massive groups, more massive galaxies have more centrally-concentrated star formation on average. Galaxies that have been identified as the centrals in their halos are marked with a black circle. In the lower row (panels $d-f$) we incorporate the galaxies with central AGN-like emission line ratios (Inc-Em sample). These galaxies are predominantly found at higher stellar mass, with low specific star formation rates, and spatially extended star formation distributions.
}\label{M_star_r50}
\end{figure*}

\begin{table*}
\begin{center}
\caption{The scale-radius ratio of galaxies split into bins of stellar mass and group halo mass. $ \langle \log(r_{50,H\alpha}/r_{50,cont}) \rangle $ is the mean log scale-radius ratio, $\sigma$ is the standard deviation of the log scale-radius ratio and $f_{cen}$ is the fraction of galaxies with $\log(r_{50,H\alpha}/r_{50,cont}) < -0.2$. }\label{r50_Mg_tab}
\begin{tabular}{rcccc}

&&$Ungrouped$&$\log\left(\frac{M_{G}}{M_{\odot}}\right) < 12.5$& $\log\left(\frac{M_{G}}{M_{\odot}}\right) \ge12.5$ \\
\hline
\multirow{3}{*}{$\log\left(\frac{M_{*}}{M_{\odot}}\right) < 10$}&$ \langle \log(r_{50,H\alpha}/r_{50,cont}) \rangle $&$-0.013\pm 0.009$&$-0.025\pm0.012$&$-0.041\pm 0.018$ \\ 
&$\sigma$&$0.093\pm0.010$&$0.086 \pm 0.009$& $0.126 \pm 0.017$ \\ 
&$f_{cen}$&$0.04^{+0.02}_{-0.02}$&$0.05^{+0.03}_{-0.02}$&$0.10^{+0.05}_{-0.04}$ \\
\hline
\multirow{3}{*}{$\log\left(\frac{M_{*}}{M_{\odot}}\right) \geq10$} &$ \langle \log(r_{50,H\alpha}/r_{50,cont}) \rangle $&$-0.005 \pm 0.010 $& $-0.063 \pm 0.017 $& $-0.124 \pm 0.020$\\ 
&$\sigma  $&$0.064 \pm 0.010$& $0.089 \pm 0.012 $& $0.122 \pm 0.011$ \\ 
&$f_{cen}$&$0.04^{+0.04}_{-0.03}$&$0.06^{+0.06}_{-0.04}$&$0.29^{+0.08}_{-0.07}$

\end{tabular}
\end{center}
\end{table*}

\subsection{The spatial extent of star formation in galaxy groups}\label{spatial_extent_in_groups_section}

We now turn to the main focus of our work, connecting the spatial extent of star formation to a galaxy's environment.  Here we will primarily focus on our star-forming sample, Final-SF, but discuss the galaxies with central AGN-like emission where appropriate (Inc-Em).

We can estimate the spatial extent of ongoing star formation using the scale-radius ratio, $r_{50,\mathrm{H}\alpha}/r_{50,cont}$ for each galaxy in our star-forming sample (see Section \ref{sec:r50}). When this ratio is large ($>1$), the current star formation is more spatially extended than the distribution of the older stars. When the ratio is small ($<1$), the star formation is less spatially extended than the distribution of older stars.

The distribution of $r_{50,\mathrm{H}\alpha}/r_{50,cont}$ vs.\ stellar mass is shown in Fig.\ \ref{M_star_r50} (upper panels) for our three different group environments.  Galaxies within our star-forming sample in the most massive groups tend to have more centrally-concentrated star formation. Across the three bins of group halo mass, the relationship between the stellar mass and the spatial extent of star formation changes. For ungrouped galaxies there is no correlation between the stellar mass and $r_{50,\mathrm{H}\alpha}/r_{50,cont}$. The lack of a correlation is consistent with the findings of \cite{Schaefer17}, where we found no dependency of the scale-radius ratio on the stellar mass of galaxies in a smaller sample. Within groups, however, this is not the case. Galaxies in groups with $M_{G}<10^{12.5} \, \mathrm{M}_{\odot}$ show a slight tendency to display smaller scale-radius ratios with increasing stellar mass. In these environments, the Spearman rank correlation coefficient between $\log(\mathrm{M_{*}/M_{\odot}})$ and $r_{50,\mathrm{H}\alpha}/r_{50,cont}$ is $\rho=-0.30$ with $p=0.005$. In groups with $M_{G}>10^{12.5} \, \mathrm{M}_{\odot}$ the strength of the correlation between the scale radius ratio and stellar mass is increased to $\rho=-0.40$, $p=0.0002$. 

In galaxy groups with halo masses above $10^{12.5} \, \mathrm{M}_{\odot}$, galaxies with stellar masses above $\sim 10^{10} \, \mathrm{M_{\odot}}$ display star formation on a smaller radial scale than for similar ungrouped galaxies. For galaxies with stellar masses less than $\sim 10^{10} \, \mathrm{M_{\odot}}$, the spatial extent of star formation appears to be less dependent on the mass of the group that the galaxy occupies. These results are summarised in Table \ref{r50_Mg_tab}, wherein we define galaxies as having `centrally-concentrated' star formation if $\log(r_{50, H\alpha}/r_{50,cont})<-0.2$. Following \cite{Schaefer17}, this threshold was chosen to be $2$ standard deviations below the mean for ungrouped galaxies. For galaxies with stellar masses greater than $10^{10}$\,M$_\odot$, the fraction of galaxies that display centrally-concentrated star formation rises from $4^{+4}_{-3} \%$ in ungrouped galaxies, to $29^{+8}_{-7} \%$ in groups more massive than $10^{12.5} \mathrm{M_{\odot}}$. For galaxies below $M_*=10^{10}$\,M$_\odot$, there is no statistically significant change in the fraction of centrally-concentrated star formers with group mass. In both ranges of galaxy stellar mass, the standard deviation from the mean of $r_{50,\mathrm{H}\alpha}/r_{50,cont}$ increases in groups with halo mass greater than $10^{12.5} \, \mathrm{M_{\odot}}$. In low and high stellar mass samples, the standard deviation in the scale-radius ratio increases by $0.033 \pm 0.020$ and $0.058 \pm 0.015$ respectively as we move to higher group mass.

In the lower panels of Fig.\ \ref{M_star_r50} we include points that have previously been excluded due to central AGN--like emission (the Inc-Em sample).  To determine the sSFR and $r_{50,\mathrm{H}\alpha}/r_{50,cont}$ for these objects we exclude H$\alpha$ flux from spaxels with emission line ratios that are inconsistent with the flux being dominated by star-forming regions.  Two main features are noticeable when we include these galaxies.  First the AGN--like objects preferentially have extended star formation and sit above the $r_{50,\mathrm{H}\alpha}/r_{50,cont}=1$ line.  This is expected as the vast majority of the AGN objects have weak LINER-like emission at their centre.  This emission is likely to be caused by post-asymptotic giant branch stars, and is visible because central star formation is no longer present. \cite{Belfiore2017} related this emission to galaxies with massive bulges that are quenching from the inside out. \cite{Spindler2018} called the central LIERs (Low-Ionisation Emission Regions) identified by \cite{Belfiore2017} `centrally-suppressed' galaxies. \cite{Spindler2018} find these preferentially at high stellar mass, but that their frequency of occurrence has no environmental dependence. This is consistent with our analysis of the rejection of galaxies with non-star-formation emission-line ratios in Section \ref{logit_regression}.

The majority of AGN objects shown in Fig.\ \ref{M_star_r50} also have low sSFR (darker red points), below $\log(sSFR/{\rm yr}^{-1})=-11$.  They are likely to be mostly quenched.  However, being quenched from inside-out, as evidenced by their spatially-extended star formation, the physical nature of the process is likely to be quite different to that of our centrally-concentrated star--forming objects.  The low sSFR and $r_{50,\mathrm{H}\alpha}/r_{50,cont}>1$ for the AGN is a consistent feature across all environments.  This points to a mass-dependent quenching mechanism being in play for these galaxies.  We find more of these galaxies in high-mass groups, but this is because high mass galaxies are preferentially found in high-mass halos.

\begin{figure}
\includegraphics{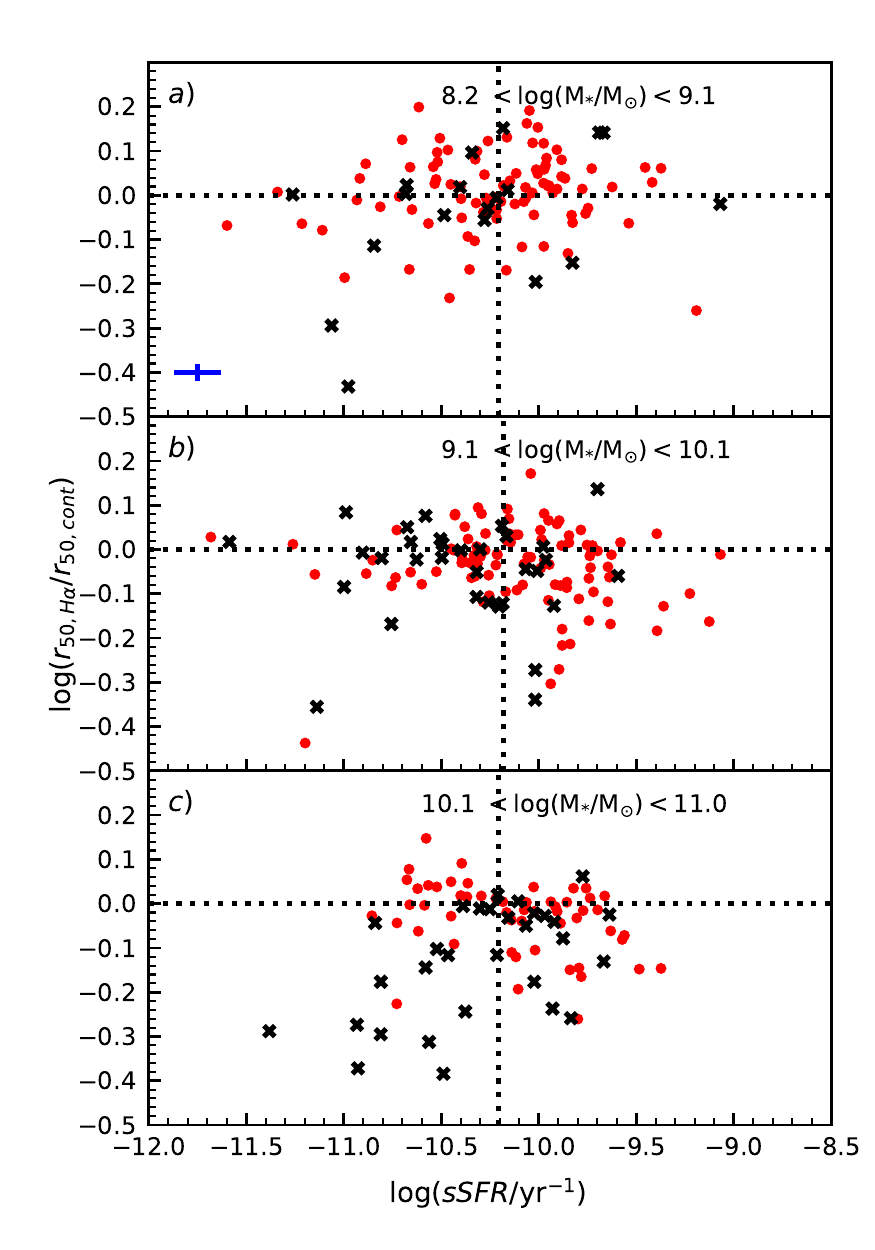}
\caption{The scale-radius ratio as a function of the sSFR of galaxies in the stellar mass intervals shown at the top of each panel. Red points represent galaxies that are either not in groups or are in groups with halo mass less than $10^{12.5} \, \mathrm{M_{\odot}}$, while black crosses are galaxies in haloes more massive than $10^{12.5} \, \mathrm{M_{\odot}}$. The vertical black dotted lines represent the mean sSFR for ungrouped galaxies in each mass bin, and the horizontal dotted line represents equally extended star formation and stellar light. Significant correlations are only seen at high stellar mass (panel c), where we find a positive correlation between sSFR and $r_{50,\mathrm{H}\alpha}/r_{50,cont}$ in high-mass groups and a negative correlation for galaxies that are in low-mass groups or are not in groups.}
\label{r50_ssfr}
\end{figure} 

A low value of $r_{50,\mathrm{H}\alpha}/r_{50,cont}$ could indicate either enhanced star formation in the centre of a galaxy or reduced star formation in the outskirts (or a combination of both).  Fig.\ \ref{M_star_r50}c suggests that the centrally-concentrated star-forming galaxies in high-mass groups have lower sSFR (points with concentrated star formation are redder), so reduced star formation in their outskirts seems more likely.  To further quantify this trend we examine the relationship between the sSFRs of galaxies with the spatial extent of star formation for different group environments in Fig.\ \ref{r50_ssfr}.  Here we have combined the galaxies that are ungrouped or in low-mass groups into one sample, to contrast them with the galaxies in high mass groups.  To highlight the qualitative differences in the quenching mechanisms at different stellar masses, we have divided our sample into three intervals of stellar mass. The various correlations for each stellar mass and environment interval are presented in Table \ref{r50_ssfr_tab}.

\begin{table*}
\begin{center}
\caption{Correlation coefficients for data displayed in Figure \ref{r50_ssfr}. $\rho$ is the Spearman rank correlation coefficient between the sSFR and the scale-radius ratio in bins of stellar mass indicated in the first column, and in bins of group halo mass indicated in the top row.}\label{r50_ssfr_tab}
\begin{tabular}{rccc}
\hline
&Ungrouped, $M_{G} < 10^{12.5}$&$M_{G} > 10^{12.5}$ & $P(\rho_{1} = \rho_{2})$\\
\hline 
$8.2<\log(\mathrm{M_{*}/M_{\odot}}) < 9.1$&$\rho=0.09$, $p=0.42$& $\rho=0.32$, $p=0.18$& $0.9 \sigma $, $p=0.19$ \\
$9.1<\log(\mathrm{M_{*}/M_{\odot}}) < 10.1$&$\rho=-0.20$, $p=0.04$& $\rho=-0.17$, $p=0.36$ & $0.17 \sigma$, $p=0.63$ \\
$10.1<\log(\mathrm{M_{*}/M_{\odot}}) < 11.0$&$\rho=-0.36$, $p=0.007$& $\rho=0.46$, $p=0.009$ & $3.7 \sigma$, $p=0.0001$\\
\hline
\end{tabular}
\end{center}
\end{table*}

For galaxies with $\mathrm{M_{*}}>10^{10.1} \, \mathrm{M_{\odot}}$ that are either ungrouped or in group halos less massive than $10^{12.5} \, \mathrm{M_{\odot}}$ (red points in Fig.\ \ref{r50_ssfr}c), there is a significant anti-correlation between the sSFR and $r_{50,\mathrm{H}\alpha}/r_{50,cont}$, with Spearman's $\rho=-0.36$ and $p=0.007$. This suggests that centrally-concentrated star formation is more often associated with increased sSFR in these intermediate/low density environments.  This may be consistent with triggered star formation due to galaxy--galaxy interactions.

For galaxies with $\mathrm{M_{*}}>10^{10.1} \, \mathrm{M_{\odot}}$ in haloes more massive than $10^{12.5} \, \mathrm{M_{\odot}}$, the distribution of star formation behaves differently (black crosses in Fig. \ref{r50_ssfr} c). For the massive group sample, the sSFR correlates positively with $r_{50,\mathrm{H}\alpha}/r_{50,cont}$. When star formation is lowered in these galaxies it is preferentially reduced at larger radius, leading to more centrally-concentrated star formation.  There is no significant correlation between sSFR and $r_{50,\mathrm{H}\alpha}/r_{50,cont}$ for lower mass galaxies in high-mass groups.

\begin{figure*}
\includegraphics{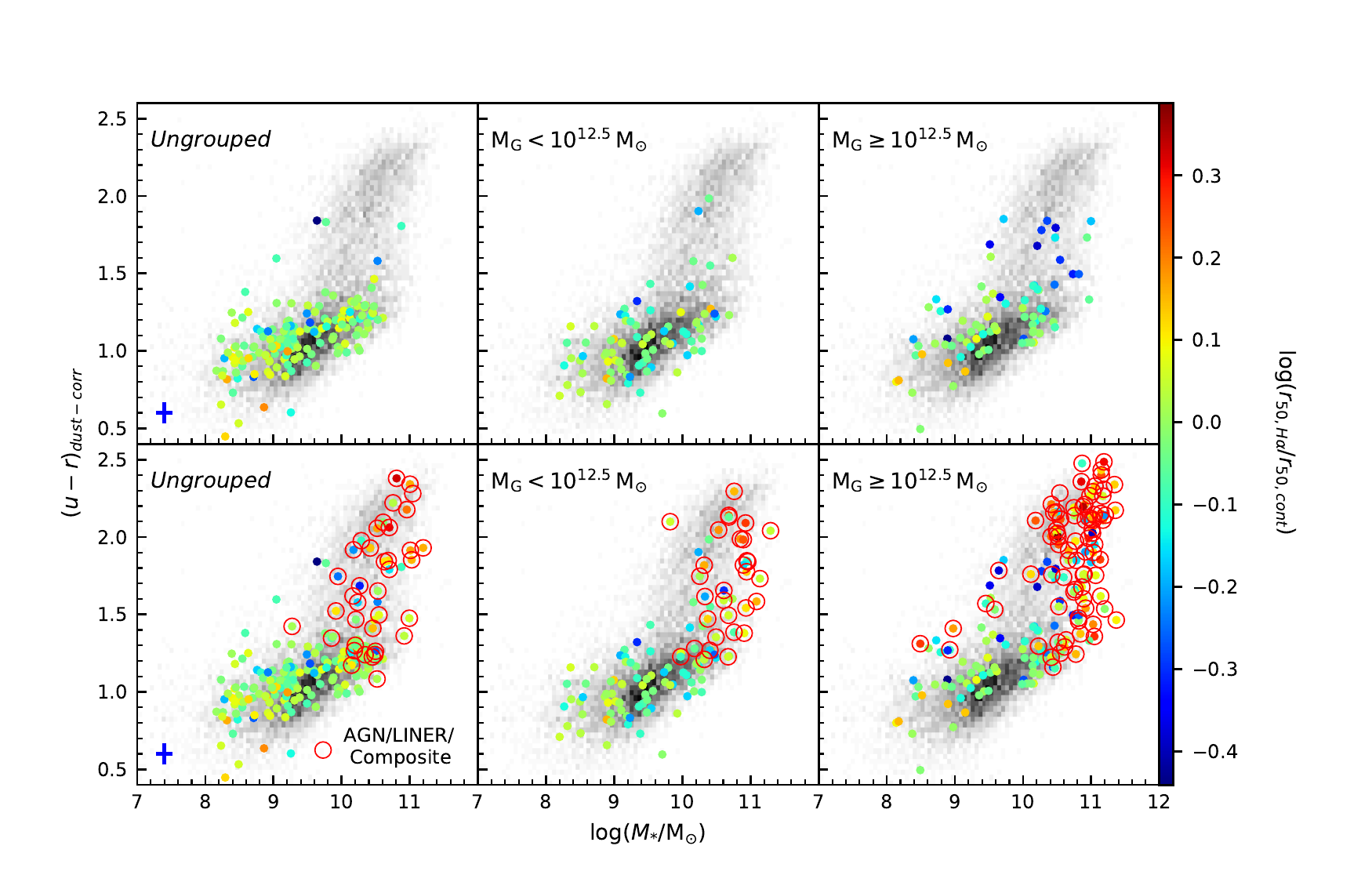}\vspace{-5mm}
\caption{The position of galaxies in our star-forming sample on the $\log(M_{*}/\mathrm{M_{\odot}})$ vs dust-corrected, rest-frame $u-r$ colour plane. Each panel shows an interval of group halo mass, The greyscale background represents the stellar mass and intrinsic $u-r$ colour of all galaxies in the GAMA survey with $z<0.1$, while the coloured points show the scale-radius ratio measured in SAMI galaxies. From left to right the panels include SAMI galaxies that are ungrouped, have measured group masses less than $10^{12.5} \, \mathrm{M_{\odot}}$ and have group masses above $10^{12.5} \, \mathrm{M_{\odot}}$. The colour scale is the same across all three group mass bins. Galaxies moving from the blue cloud to the red sequence in massive groups tend to have centrally-concentrated star formation. The upper panels shows only our star--forming sample (Final-SF), while the lower panels also includes galaxies with central AGN-like emission (red circles).}\label{colour_magnitude}
\end{figure*}

\subsubsection{The central star formation density in massive groups}
We have calculated the star formation rate surface density in a central $1\arcsec$ aperture in each of the star-forming galaxies in our sample. We have used these values as a further test of whether the central enhancement of star formation is able to explain the reduction in $r_{50,\mathrm{H}\alpha}/r_{50,cont}$ in high-mass groups. For galaxies with stellar masses in the range $10^{10}<M_{*}/\mathrm{M_{\odot}}<10^{10.5}$, we compare the average central star formation rate surface density for normal galaxies and those with centrally-concentrated star formation. This limited range of stellar mass was chosen to minimise the influence of the relationship between central star-formation rate surface density and stellar mass. We find that in groups with halo mass greater than $10^{12.5} \, \mathrm{M_{\odot}}$ the mean central star formation rate surface density is $\log(\Sigma_{sfr}/\mathrm{M_{\odot} \, yr^{-1} \, kpc^{-2}})=-1.41 \pm 0.11$ and $\log(\Sigma_{sfr}/\mathrm{M_{\odot} \, yr^{-1} \, kpc^{-2}})=-1.20 \pm 0.19$ for normal and centrally-concentrated star-formers respectively. This yields a difference of $0.21 \pm 0.22$, which is below one sigma significance. The central star formation rate surface density in galaxies with centrally-concentrated star formation is not significantly different from other galaxies with similar stellar mass in the same environments. 
The consistent central star formation surface densities also imply that the difference in concentration cannot be caused by the profiles drop uniformly and the outer parts of the profile dropping below our detection limit.

\subsubsection{The spatial extent of star formation and the colour-mass diagram}
The position of a galaxy on the colour--mass diagram is commonly used to diagnose its current evolutionary state. Galaxies in the blue cloud are usually star-forming or have had a recent burst of star formation, while galaxies in the red sequence are often passive and have no evidence of star formation within the last several billion years. Galaxies in between the blue cloud and red sequence are often considered to be in the process of being quenched or having their star formation rejuvenated. We display the locations of the galaxies from our Final-SF sample in this parameter space in Fig.\ \ref{colour_magnitude} (top panels), dividing the sample into bins of group mass.  The colours are dust--corrected and shifted to the rest--frame of the galaxies. For our star--forming sample, we see a larger fraction of galaxies with intermediate $u-r$ colours in high mass groups.  Of these galaxies at intermediate colours, many have centrally-concentrated H$\alpha$ emission.  There is no evidence of such a trend in the ungrouped or low mass group sample.  This provides further evidence that the galaxies with centrally concentrated star formation in high mass groups are in the process of quenching.

In the lower panels of Fig.\ \ref{colour_magnitude} we show the location of galaxies in the colour--mass plane, including those that have central AGN--like emission (marked by a red circle). The majority of them are on the red sequence, or at intermediate colours. This reiterates the point made above that most of the galaxies with central AGN-like emission are quenched or quenching. The distribution of AGN in colour--mass space appears the same for our three environmental bins, with the exception of there being a larger number of the most massive galaxies in high--mass groups.  

\subsection{Satellite and central galaxies}
Galaxies that are environmentally quenched may experience different processes that shut down their star formation depending on where they sit within their parent group halo. A galaxy that is at the centre of a group will be less likely to experience, for example, ram pressure stripping than a galaxy that is its satellite. In galaxy groups that are dynamically relaxed, the most massive object will tend to sit at the centre of the halo, but this is not always the case \citep[e.g.][]{Skibba2011,Olivia-Altamirano2014}. The group catalogue of \cite{Robotham2011} calculates an iterative group centre that is robust to the effects of massive galaxies in falling into groups. In this process, the galaxy that is most distant from the centre of light of the group is rejected and the centre of light is recalculated. This process is repeated until two galaxies remain, at which point the brightest of that pair is decided as the iterative centre. \cite{Robotham2011} reports that in $95$ per cent of cases the iterative centre is the same as the brightest group galaxy.  We term galaxies in our sample that are identified as the iterative centre of their group as `centrals' and all other group galaxies as `satellites'.  The centrals are marked by a black circle in Fig.\ \ref{M_star_r50}a-c.

As would be expected, the number of centrals we observe in groups less massive than $10^{12.5} \, \mathrm{M_{\odot}}$ is large, representing a total of $48$ of the $85$ star-forming galaxies in this group mass bin (most of these groups have low multiplicity). These galaxies preferentially occupy the higher end of the mass distribution. We can measure the correlation between stellar mass and $r_{50,\mathrm{H}\alpha}/r_{50,cont}$ separately for the centrals and satellites, but find very little difference between the two populations. For satellites, the Spearman rank correlation coefficient is $\rho = -0.290$, with $p=0.068$, while for centrals it is $\rho=-0.287$, with $p=0.058$. The correlation between the spatial extent of star formation and stellar mass is similar for the satellites and centrals of groups less massive than $10^{12.5} \, \mathrm{M_{\odot}}$. 

Galaxies that are the centrals of massive groups are by definition rare and, having high stellar mass, are often already passive. Of the $82$ star-forming galaxies in groups more massive than $10^{12.5} \, \mathrm{M_{\odot}}$, only $12$ are centrals. This low number means that we are unable to make definitive statements about whether the spatial distribution of star formation in centrals differs in comparison to the star formation distribution in satellite galaxies in massive groups. In Figure \ref{M_star_r50} we have marked central galaxies with a black circle. The correlation coefficient between $\log(M_{*}/\mathrm{M_{\odot}})$ and $\log(r_{50,\mathrm{H}\alpha}/r_{50,cont})$ goes from $\rho=0.40$, $p=0.0002$ with centrals included, to $\rho=0.41$, $p=0.0005$ for satellites only.

In Figure \ref{r50_ssfr}, we showed that the scale-radius ratio in galaxies with stellar masses greater than $10^{10.1} \, \mathrm{M_{\odot}}$ in groups with halo mass greater than $10^{12.5} \, \mathrm{M_{\odot}}$ is correlated with their sSFRs (black crosses in Fig.\ \ref{r50_ssfr}c).  If we remove centrals from this sample, the strength of the correlation increases slightly, from $\rho=0.46$ and $p=0.009$ to $\rho=0.53$ and $p=0.006$, though the sample size is reduced to just $26$ galaxies. In all other analysis, unless otherwise stated we do not make any distinction between satellite and central galaxies.

\subsection{Galaxy-galaxy tidal interactions}
\begin{figure}
\includegraphics{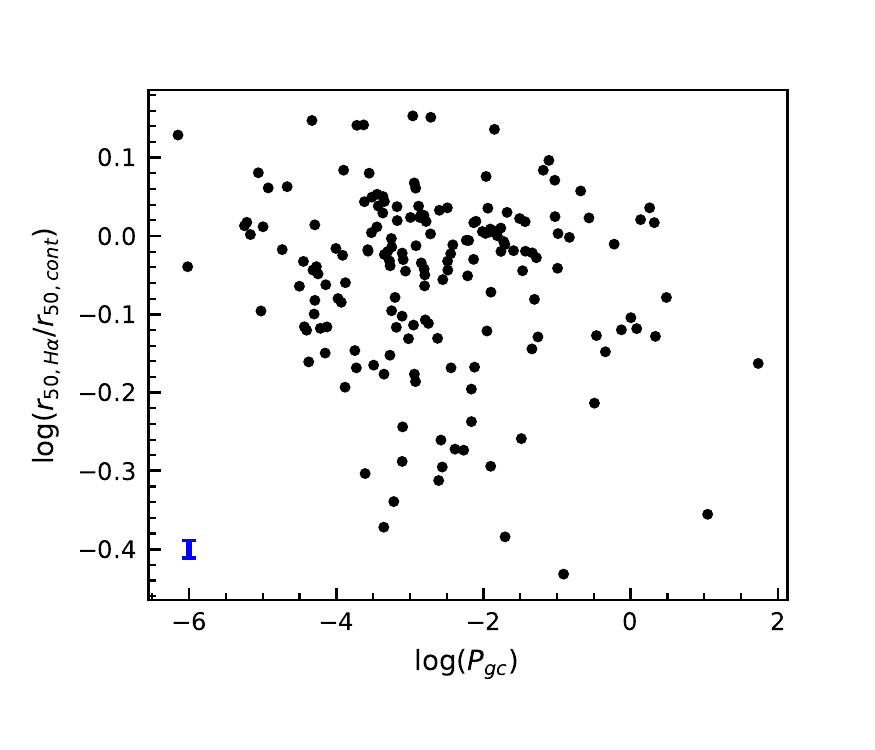}
\caption{The scale-radius ratio as a function of the tidal perturbation parameter for galaxies in groups. With a Spearman rank correlation coefficient of $\rho=-0.06$ with $p=0.41$, there is no evidence that the current tidal perturbation influences the radial distribution of star formation in these galaxies.}\label{tid_pert_r50}
\end{figure}

A number of authors have suggested that dynamical disturbances driven by tidal interactions between galaxies in groups can cause gas to fall towards the centres of galaxies \citep[e.g.][]{Hernquist89,Moreno2015}. While this may cause the enhancement of star formation on short timescales, the consumption of gas by star formation induced by the interaction can ultimately cause the galaxy to become quenched earlier than in ungrouped galaxies. We can estimate the strength of the current tidal interaction using the perturbation parameter of \cite{Dahari84} and \cite{Byrd90},

\begin{equation} \label{tid_pert_eqn}
P_{gc}=\left(\frac{M_{c}}{M_g}\right) \times \left(\frac{r_{g}}{d_{gc}}\right)^{3},
\end{equation}
where $M_{c}$ is the mass of the companion, $M_{g}$ is the mass of the galaxy being perturbed, $r_{g}$ is the optical size of the galaxy and $d_{gc}$ is the projected distance between the galaxy and its companion. \cite{Dahari84} and \cite{Byrd90} defined this $r_{g}$ in terms of the optical sizes of galaxies as measured by hand from photographic plates, so we shall approximate this with the $r$-band $r_{90}$, the radius that contains $90 \%$ of the flux, as given by the GAMA S{\'e}rsic photometric fits. \cite{Byrd90} showed that gas infall is expected when the perturbation parameter is greater than $\sim0.01-0.1$, depending on the halo-to-disc mass ratio. We have calculated the perturbation parameter for all possible pairs of galaxies in the GAMA Galaxy Group Catalogue, and use the greatest perturbation within a group to estimate the tidal effects on a galaxy. Using this estimator of the strength of the tidal forces experienced by each galaxy in the group we find no evidence that tidal interactions change the radial extent of star formation in galaxy groups. This is shown in Figure \ref{tid_pert_r50}. We must note that this does not necessarily imply that tidal interactions have no effect on the distribution of star formation in galaxies because there could be a considerable time delay between a tidal interaction and the movement of gas to the centres of the galaxies.

\subsection{Tidal interactions with the group potential}
\begin{figure}
\includegraphics{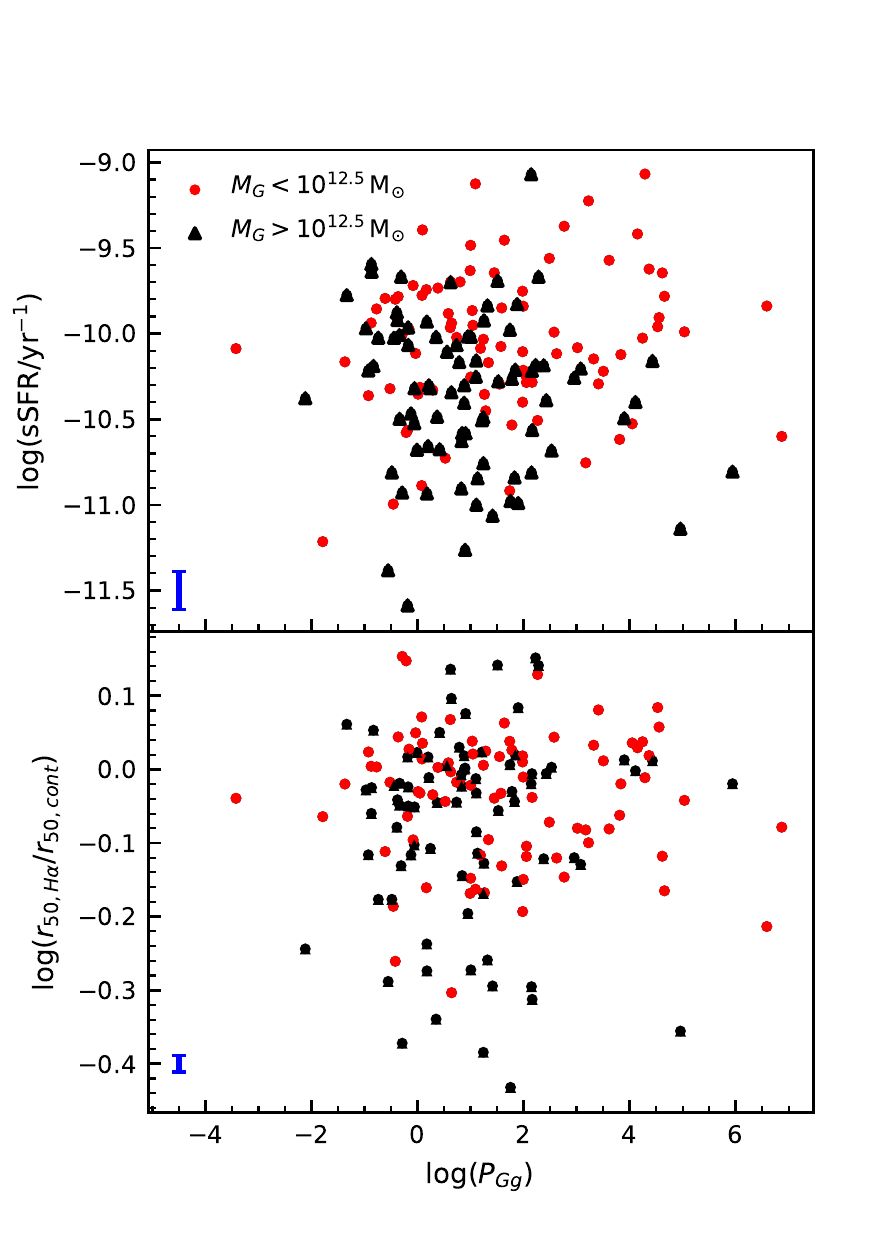}
\caption{The star-forming properties of galaxies compared to the tidal influence of the group halo. In the upper and lower panels we compare respectively the sSFR and scale radius ratio to the group tidal perturbation parameter, $P_{Gg}$. Black points denote galaxies in groups more massive than $10^{12.5} \, \mathrm{M_{\odot}}$ and red points indicate galaxies in lower mass groups. There is no relationship with either measurement and $P_{Gg}$, though a systematic trend with the sSFR and $r_{50,\mathrm{H}\alpha}/r_{50,cont}$ does exist, both quantities being lower on average in more massive groups.}\label{g_tid_pert_all}
\end{figure}

The galaxy groups in this sample can be as massive as $10^{14} \, \mathrm{M_{\odot}}$. This means that tidal perturbations in individual galaxies can be caused by their gravitational interaction with the group potential as a whole. For a galaxy with mass $M_{g}$ with radius $r_{g}$ located at a distance $d$ from the centre of a group halo of mass $M_{G}$, the tidal perturbation is given by
\begin{equation}
P_{Gg}=\left( \frac{M_{G}}{M_{g}}\right) \times \left(\frac{r_{g}}{d}\right)^{3},
\end{equation}
following \cite{Boselli2006}.
We calculate this value using the group mass and group-centric distances provided by the GAMA Galaxy Group Catalogue, and compare it to various properties of each galaxy. In Figure \ref{g_tid_pert_all} we compare the group tidal perturbation to the star formation rates and distributions in our sample. We find no correlation between the group tidal perturbation and the star-forming properties of the galaxies. The lack of correlation does not necessarily rule out that tidal interactions with the group potential could alter the properties of galaxies because the metric for tidal perturbation used has large systematic uncertainties. Each galaxy's distance $d$ from the centre of the group is subject to projection effects, in addition to the group centre being poorly defined for groups with only a few members. The use of stellar masses for the mass of each galaxy is also systematically low; the amount of dark matter in each galaxy is unknown but likely higher than the stellar mass.

\subsection{Projected phase space} \label{Projected_phase_space}
Some recent works have made use of projected phase space diagrams as a means of diagnosing particular processes that may be acting on galaxies in clusters \citep[e.g.][]{Oman2013,Jaffe15,Oman2016}. In this scheme galaxies are placed in phase space with position and velocity measured relative to the host halo. Galaxies with velocities greater than the group velocity dispersion and within $0.5 \, R_{200}$ are likely to undergo ram pressure stripping, low velocity galaxies far from the group centre are likely to be on their first passage through the group, and slow-moving galaxies close to the group centre are likely virialised within the group. 

We calculate $R_{200}$ for the groups in our sample using the prescription of \cite{Finn2005},
\begin{equation}
R_{200}=1.73 \frac{\sigma_{v}}{1000 \, \mathrm{km \, s^{-1}}} \frac{1}{\sqrt{\Omega_{0} + \Omega_{\Lambda}(1+z)^{3}}} h^{-1}_{100} \, \mathrm{Mpc},
\end{equation}
where $\sigma_{v}$ is the group velocity dispersion and $z$ is the systemic redshift of the group.

\begin{figure}
\includegraphics{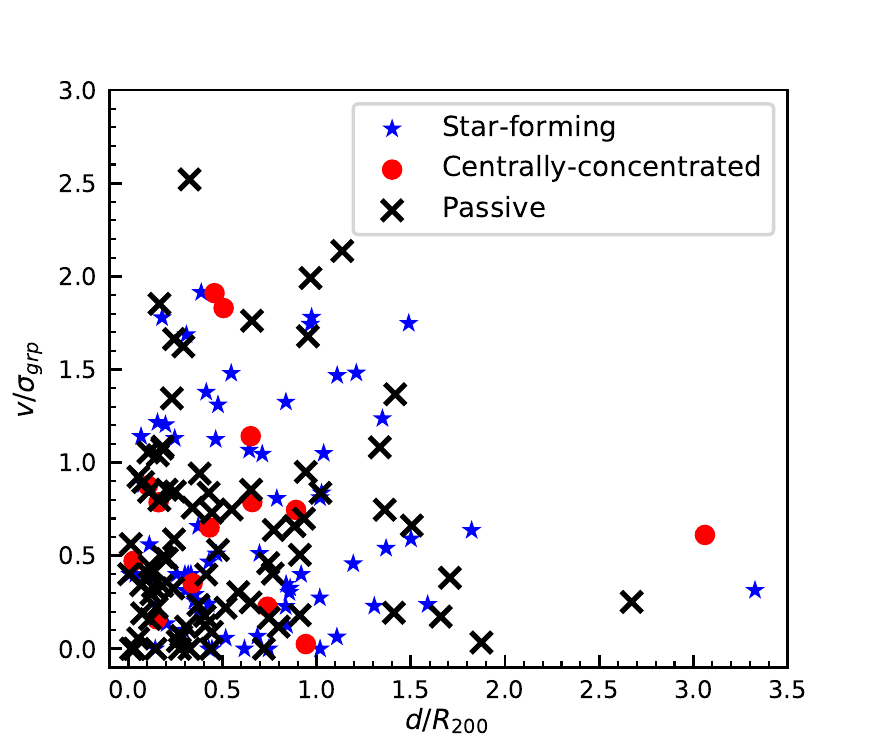}
\caption{Projected phase space diagram for galaxies in groups with $M_{G}>10^{12.5} \, \mathrm{M_{\odot}}$. The horizontal axis is the distance each galaxy is from the centre of the group, in units of $R_{200}$. The vertical axis is the velocity of each galaxy relative to the systemic velocity of the group normalized by the velocity dispersion of the group. Blue stars and red circles are star-forming galaxies, and black crosses are passive galaxies. Red circles represent star-forming galaxies with centrally-concentrated star formation.}\label{PPS_cen_con}
\end{figure}

In Figure \ref{PPS_cen_con} we display all  galaxies in groups more massive than $10^{12.5} \, \mathrm{M_{\odot}}$ in projected phase space. Galaxies showing centrally-concentrated star formation ($\log(r_{50, H\alpha}/r_{50,cont})<-0.2$) follow the same distribution in projected phase space as other star-forming galaxies. A two-sample Kolmogorov-Smirnov test comparing the centrally-concentrated star formers to the normal star-forming galaxies along each dimension of projected phase-space showed that the distributions are not significantly different. In group-centric radius the K-S statistic is $D=0.21$ with p-value $0.62$, and in relative velocity the K-S statistic is $D=0.24$ with p-value $0.46$.  All but one of the galaxies with centrally-concentrated star formation exist within $R_{200}$ of their respective groups, and most have line-of-sight velocities relative to the group less than the group velocity dispersion.

\subsection{Distribution of star formation vs group-centric radius}
In Section \ref{Projected_phase_space} we saw that the distribution of centrally-concentrated star-forming galaxies within $R_{200}$ of groups more massive than $10^{12.5} \, \mathrm{M_{\odot}}$ is not significantly different from other star-forming systems in projected phase space. We find no strong trend between $r_{50, H\alpha}/r_{50,cont}$ and group-centric radius for galaxies within groups. To highlight the difference between the galaxy populations within and outside of galaxy groups, we have matched galaxies from the ungrouped sample to the nearest group halo, within $\pm 1000 \, \mathrm{km} \, \mathrm{s^{-1}}$ from the matched group systemic velocity. 

\begin{figure}
\includegraphics{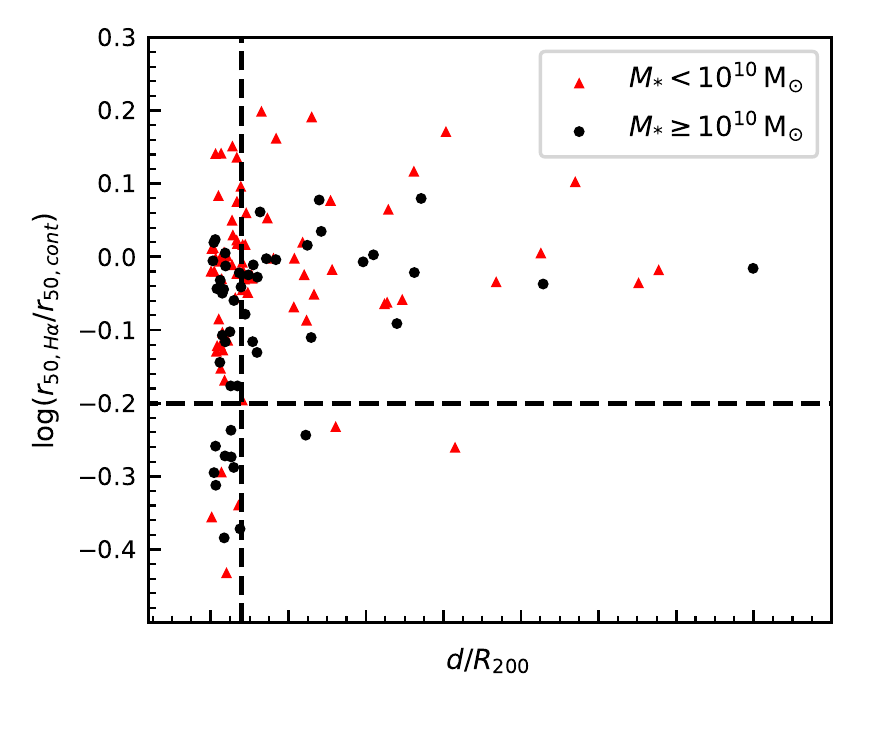}
\caption{The scale radius ratio for galaxies as a function of distance from the centre of the nearest group. This sample includes both galaxies that are in the GAMA Galaxy Group Catalogue and galaxies whose nearest group halo is more massive than $10^{12.5} \, \mathrm{M_{\odot}}$. The horizontal dashed line is the fiducial $\log(r_{50, H\alpha}/r_{50,cont}) =-0.2$ dividing line below which we say a galaxy has centrally-concentrated star formation. For $M_{*}>10^{10} \, \mathrm{M_{\odot}}$, galaxies with centrally-concentrated star formation exist almost exclusively within $R_{200}$, though there is no correlation with projected group-centric distance within $R_{200}$ in these groups.}\label{rg_r200_r50}
\end{figure}

Fig.\ \ref{rg_r200_r50} shows the scale-radius ratio in galaxies in and around groups more massive than $10^{12.5} \, \mathrm{M_{\odot}}$.  We have split the sample into two stellar mass intervals, above and below $M_{*}=10^{10} \, \mathrm{M_{\odot}}$.  Within these groups, closer than $R_{200}$ to the centre of the group, there is no correlation between distance, $d$, and $r_{50, H\alpha}/r_{50,cont}$. For the higher stellar mass sample  
within $R_{200}$ the fraction of star-forming galaxies with centrally-concentrated star formation is $35^{+9}_{-8}$ per cent. Outside of $R_{200}$ this fraction drops to $7^{+7}_{-4}$ per cent\footnote{These fractions differ slightly, but not significantly from the fractions presented in subsection \ref{spatial_extent_in_groups_section}.  This is because those fractions included galaxies that are associated with a group, but may have been further than $R_{200}$ from the group centre.}. There are relatively few galaxies with $M_{*}>10^{10} \, \mathrm{M_{\odot}}$ outside of $R_{200}$ that show the signatures of centrally confined star formation.  Existing within $R_{200}$ of a group with halo mass greater than $10^{12.5} \, \mathrm{M_{\odot}}$ appears to be the primary factor in determining the outside-in quenching of star formation for galaxies with $M_{*} > 10^{10} \, \mathrm{M_{\odot}}$.
For galaxies less massive than $M_{*}=10^{10} \, \mathrm{M_{\odot}}$ there is no radial trend in the fraction of galaxies with concentrated star formation.

\subsection{Nearest neighbour interactions}
Data from large-scale surveys has suggested that interactions between galaxies and their nearest neighbours may be able to either enhance or suppress star formation \citep[e.g.][]{Patton2013, Davies2015}. To test this with SAMI, we use nearest-neighbour distances from the GAMA spectroscopic catalogue, including galaxies that are not necessarily in the group catalogue, but are more luminous than $\mathrm{M}_r = -18.5$ mag. 
For the sample as a whole, we find no dependence of the sSFRs of galaxies on the distance to the nearest neighbour, though we note that our sample size is several orders of magnitude smaller than the large samples acquired through single-fibre spectroscopy \citep[e.g.][]{Ellison2008}. The ratio $r_{50, H\alpha}/r_{50,cont}$ appears to be reduced at small separations between galaxies, but the nearest-neighbour distance is strongly correlated with group mass, obscuring any strong conclusion. To resolve this issue we performed a partial Spearman rank correlation analysis between the group mass, nearest-neighbour distance, and $r_{50, H\alpha}/r_{50,cont}$ ratio for galaxies with group mass estimates. When the group mass is taken into account, the Spearman rank correlation coefficient between nearest neighbour distance and $r_{50, H\alpha}/r_{50,cont}$ is found to be small and not significant ($\rho = 0.09, \, p=0.26$).

\subsubsection{The effect of close pairs}
As a further test for the effect of close pairs in galaxy groups we remove all galaxies that have a nearest neighbour of any luminosity within $50 \, \mathrm{kpc}$ (projected distance) in the GAMA close pair catalogue. With the close pairs removed from our sample we repeat the analysis of Figs.\ \ref{M_star_r50} and \ref{r50_ssfr}. Removing the galaxies in close pairs has the greatest impact on the high-mass group sample, with the number of galaxies dropping from $82$ to $53$. In the massive groups and for galaxies more massive than $10^{10.1} \, \mathrm{M_{\odot}}$ the Spearman rank correlation is reduced from from $\rho=0.46,\, p=0.009$ to $\rho=0.32,\, p=0.19$.   However, the difference between the correlation coefficients is not significant (the probability of them being different is 0.40).  These results as well as those for other mass intervals are given in Table \ref{r50_ssfr_no_pairs}.  We do not measure any significant difference when close pairs are removed from any of the samples.

\begin{table*}
\begin{center}
\caption{The strengths of the correlations between the sSFR and scale-radius ratio in galaxies in different bins of stellar mass and group halo mass. Also shown are the strengths of the correlations when close pairs ($<50 \, \mathrm{kpc}$ separation) are removed. We have provided the significance of the difference in correlations, $\sigma$, and the probability $P$ that they are the same. In our sample, galaxies in close pairs do not significantly affect our results.}\label{r50_ssfr_no_pairs}
\begin{tabular}{rcllrc}
&&Ungrouped, $M_{G} < 10^{12.5}$& $P(\rho_{pairs} = \rho_{no \, pairs})$&$M_{G} > 10^{12.5}$ & $P(\rho_{pairs} = \rho_{no \, pairs})$\\
\hline 
\multirow{2}{*}{$8.2<\log(\mathrm{M_{*}/M_{\odot}}) < 9.1$}&pairs&$\rho=0.09$, $p=0.42$& $0.28\sigma$&$\rho=0.32$, $p=0.18$ &$-0.05\sigma$ \\
&no pairs&$\rho=0.13$, $p=0.26$&$P=0.76$&$\rho=0.30$, $p=0.26$& $P=0.96$\\
\hline

\multirow{2}{*}{$9.1<\log(\mathrm{M_{*}/M_{\odot}}) < 10.1$}&pairs&$\rho=-0.20$, $p=0.04$&$6\times10^{-4}\sigma$& $\rho=-0.17$, $p=0.36$ & $0.49 \sigma$ \\
&no pairs&$\rho=-0.20$, $p=0.06$&$P=1.0$&$\rho=-0.02$, $p=0.94$&$P=0.62$\\
\hline

\multirow{2}{*}{$10.1<\log(\mathrm{M_{*}/M_{\odot}}) < 11.0$}&pairs&$\rho=-0.36$, $p=0.007$&$0.04 \sigma$ &$\rho=0.46$, $p=0.009$ & $-0.53 \sigma$\\
&no pairs&$\rho=-0.35$, $p=0.016$&$P=0.96$&$\rho=0.32$, $p=0.19$& $P=0.60$

\end{tabular}
\end{center}
\end{table*}

\cite{Davies2015} reported different environmental effects for galaxies depending on the relative mass of their nearest neighbour. Galaxies with a more massive companion are more likely to be quenched, while those with a less massive companion are more likely to have their star formation enhanced. Our current same size is not large enough to make any meaningful comment about this distinction.

\section{Discussion}\label{Discussion}
Taking advantage of the SAMI Galaxy Survey and the GAMA Galaxy Group Catalogue we have examined how the spatial extent of star formation (as parameterized by $r_{50, H\alpha}/r_{50,cont}$) and the integrated specific star formation rate (sSFR) depend on environment.  The rationale behind this is that the possible mechanisms at work may modify these measured quantities in different ways.

The SAMI data show that both $r_{50, H\alpha}/r_{50,cont}$ and sSFR are dependent on the mass of the host halo.  In particular, the signatures are different for galaxies with group masses greater than and less than $M_{G} = 10^{12.5}$\,M$_\odot$.  We reiterate that the mean group mass of our high mass groups is $10^{13.2}$\,M$_\odot$ (see Fig.\ref{G_mass_hist}), so we are considering halos in the group regime (not clusters).  We discuss the high and low group mass regimes in turn below.

\subsection{Massive groups, $M_{G} > 10^{12.5} \, \mathrm{M_{\odot}}$}
We find that star-forming galaxies (defined as having $\ewha >1$\,\AA) in massive groups have lower sSFRs than ungrouped galaxies or galaxies in low mass groups. Similar trends have been seen by previous studies, but the reduction of the star formation rates in star-forming galaxies with environment has been the subject of some debate in the literature. \cite{Wetzel2012} found no change in the peak sSFRs of star-forming galaxies as a function of environment, and concluded that galaxies must transition from star formation to quiescence rapidly in dense environments. This is in agreement with many others such as \cite{Balogh04} and \cite{Peng2010}. Similarly, \cite{Paccagnella2016} found no change in parameters of the star formation rate main sequence in galaxy clusters, but did report a larger number of galaxies in transition between star formation and quiescence at small clustercentric radii. This echoes the reduction in the average star formation rates of star-forming galaxies with clustercentric radius seen by \cite{vonderLinden2010}. 

At the heart of this discrepancy in the literature is the definition of what constitutes a `star-forming' galaxy. Studies that employ a deeper lower limit on the star formation rate of star-forming galaxies tend to see a trend with environment. The impact of the cutoff used to separate the star-forming and passive populations of galaxies is discussed at length by \cite{Taylor2015} in the context of defining the blue cloud and red sequence on the colour-mass diagram. The same arguments apply here. Our definition of `star-forming' is primarily constrained by the detectability of H$\alpha$ emission within the galaxies in our sample. Since the SAMI spectroscopy is relatively deep, and since the integral field data are less affected by aperture effects than single-fibre spectroscopy, we are able to detect lower levels of star formation in galaxies as they make the transition towards quiescence.

Our Fig.\ \ref{SF_MS_Gmass_hist} shows that the peak sSFR is shifted to lower values in our high mass groups.    In this analysis we have used all galaxies (Full-EM sample), including those with AGN-like central spectra (only using star forming spaxels), but because of the very small number of AGN-like objects at low/intermediate stellar mass  this result is robust against issues related to bias caused by AGN.  The shift in the mean sSFR we measure is between $\Delta\log(sSFR/{\rm yr}^{-1})=-0.38\pm0.10$ and $-0.64\pm0.14$ (from low to high stellar mass) and a similar result is found if we estimate the median.  This is a significant difference in all three galaxy stellar mass intervals.  The mean over all galaxy stellar masses is $-0.45\pm0.07$, and there is no significant change in this value with stellar mass.  \cite{Rasmussen12} uses UV photometry from GALEX to find a similar decrease in the sSFR of low mass star forming galaxies in groups.  However, they find no suppression for high stellar mass galaxies ($M_*>10^{10}$\,M$_\odot$).  This difference may be caused by the more restrictive dynamic range used to define star--forming galaxies ($\log(sSFR/yr^{-1})>-10.5$) in the work of \cite{Rasmussen12}.  In contrast, \cite{Wetzel2012} (fitting a skewed Gaussian to the star-forming and passive populations) find differences of $\Delta\log(sSFR/{\rm yr}^{-1})=-0.2$ or less (their Fig. 2) and argue there is no significant difference between centrals and satellites that are star forming, whatever the mass of the halo.  Potential causes of the difference with \cite{Wetzel2012} include aperture effects and their use of D$_{n}$4000 as a star formation indicator when H$\alpha$ is weak or obscured by AGN, as this provides an upper limit to the true sSFR.

Using a sample of purely star forming galaxies (Final-SF) we find that galaxies in groups with $M_{G}>10^{12.5} \, \mathrm{M}_{\odot}$, with stellar masses greater than $\sim 10^{10} \, \mathrm{M}_{\odot}$, have more centrally-concentrated star formation than the same mass galaxies in lower mass halos.  This trend is not observed for galaxies at lower stellar masses.  Such a result is in agreement with our previous analysis measuring the radial H$\alpha$ gradients of SAMI galaxies as a function of 5th-nearest neighbour density \citep{Schaefer17}.  The process that is environmentally quenching these galaxies is acting from outside--in.

While the observations above do not tell us directly what the process of quenching is, the measurements of sSFR and concentration have at least two implications for our understanding of the quenching process: 
\begin{enumerate}
\item
the lower sSFR of star-forming galaxies in high-mass groups means that the environmental quenching of star formation within these groups cannot be faster than the time-scale taken from the birth to death of OB stars.  It is these OB stars that excite H$\alpha$ emission in HII regions. If quenching occured on a timescale shorter than this, we would not see the lowering of the mean sSFR in the most massive groups. Instead we would only see an increase in the quenched fraction of galaxies.

\item
Either low-mass galaxies and high-mass galaxies quench by a different mechanism, or a single mechanism causes the effects that we observe, but this mechanism affects the distribution of star formation in a way that depends on the stellar mass. For example, if the timescale for the outside-in quenching of low-mass galaxies is substantially shorter than for high-mass galaxies, we would observe fewer centrally-concentrated star-forming galaxies at low stellar mass.
\end{enumerate}

For our pure star-forming sample (Final-SF) we have also quantified the change in the scatter of $r_{50,\mathrm{H}\alpha}/r_{50,cont}$ with environment (see table \ref{r50_Mg_tab}).  There is an increased scatter at high group mass and for high mass galaxies this increased scatter is preferentially to lower $r_{50,\mathrm{H}\alpha}/r_{50,cont}$. That is, while some galaxies in more massive groups have radial star-formation distributions similar to those in ungrouped galaxies, a significant fraction have more centrally-concentrated star formation ($\log(r_{50,\mathrm{H}\alpha}/r_{50,cont}) < -0.2$), and very few have more extended star formation. In particular,  in our final star-forming sample, $29^{+8}_{-7}$ per cent of galaxies more massive than $M_{*}=10^{10} \, \mathrm{M_{\odot}}$ in the massive group sample have $\log(r_{50,\mathrm{H}\alpha}/r_{50,cont}) < -0.2$, compared to $4^{+2}_{-2}$ per cent in ungrouped galaxies. The increase in the scatter in $r_{50,\mathrm{H}\alpha}/r_{50,cont}$ suggests some inefficiency in the outside-in quenching process.  It is not clear whether this process acts on all galaxies that fall into these groups, or whether the inclination of the galaxy disc to the direction of passage through the intergalactic medium influences the change in the star-formation distribution, as has been predicted by simulations of ram pressure stripping in galaxy groups and clusters \citep[e.g.][]{Bekki2014}.

Figure \ref{colour_magnitude} suggests that the star-forming galaxies in our sample occupy different distributions in the colour-mass diagram depending on the mass of the parent halo. Above $M_{G} = 10^{12.5} \, \mathrm{M_{\odot}}$, a larger fraction of galaxies is in transition between the blue cloud and the red sequence, particularly for systems with $\mathrm{M}_{*}>10^{10} \, \mathrm{M_{\odot}}$. This has been reported by a number of studies using unresolved data in the past \citep[e.g.][]{Hogg2004,Balogh04,Bassett2013}. \cite{Schawinski2014} showed that late type galaxies in halos more massive than $10^{12} \mathrm{M_{\odot}}$ are much more likely to lie off the blue cloud in the colour-mass diagram. Their modeling showed that the timescale for a late-type galaxy to make the transition from the blue cloud to the red sequence is of order several Gyr, but that this can be accelerated by various environmental processes. Our results indicate that a fraction of this transitioning population of galaxies can be explained by an outside-in quenching mechanism that is consistent with gas stripping.

Within groups with halo mass greater than $10^{12.5} \, \mathrm{M_{\odot}}$, we find no significant projected phase--space differences between galaxies with centrally-concentrated star formation and passive or normally star--forming galaxies.
If outside--in quenching of star formation started in these galaxies soon after their infall into these massive groups, we would expect a greater separation of these two types of galaxies in projected phase-space. However, if the efficiency of the mechanism that causes this change in the star formation morphology is poor, or its onset is delayed, the distribution of galaxies with centrally-concentrated star formation is easily explained.  Being within $\sim R_{200}$ is important to the outside--in quenching to take place, as the fraction of centrally concentrated star--forming galaxies is significantly lower outside of this radius.

\cite{Jaffe15} detected the $21 \, \mathrm{cm}$ neutral hydrogen emission line in galaxies outside the virialised region of galaxy clusters, but did not detect this line for galaxies inside the virialised region, where they have presumably resided in the clusters for several cluster dynamical times. Given the difference in halo mass between the groups in our sample and the cluster studied by \cite{Jaffe15} it is difficult to draw direct comparisons, but it seems probable that many of the galaxies with centrally-concentrated star formation are not on their first passage into their host groups.  This is consistent with \cite{Brown2017} who find a deficit of neutral gas, even in relatively low mass groups ($10^{12} - 10^{13.5}$\,M$_\odot$).

Comparing the distribution of these galaxies to the simulations of projected phase-space performed by \cite{Oman2013}, we can conclude that the majority of centrally-concentrated star-forming galaxies have been in their groups for perhaps over three Gyr \citep[see][their Fig.\ 4]{Oman2013}. We interpret the distribution of these galaxies in projected phase--space as a sign that the quenching of star formation by this outside-in mechanism is not instantaneous and persists over several group-crossing times.

\cite{Peng15} used stellar metallicities to infer that starvation is expected to be the primary quenching route.  The single fibre SDSS data used by \cite{Peng15} only samples the inner parts of galaxies, meaning that a possible picture in high mass groups is for galaxies to suffer partial outside-in ram pressure stripping.  The central parts of the galaxies, that have not been stripped, then slowly quench via starvation.

\subsection{Low-mass groups, $M_{G} < 10^{12.5} \, \mathrm{M_{\odot}}$}
In contrast to the environmental suppression of star formation in galaxies in the high-mass group sample, we find that for galaxies in low-mass groups (that is, with halo mass $M_{G} < 10^{12.5} \, \mathrm{M_{\odot}}$), there is little evidence for environmental quenching in galaxies. We find that in these systems there is some evidence that centrally concentrated star formation is related to an {\it increase} in sSFR for galaxies with stellar mass $>10^{10.1}$\,M$_\odot$.

A similar effect was noted by \cite{Davies2015} when studying the star formation rates of pairs of galaxies in the GAMA catalogue. They showed that the more massive galaxies in pairs have centrally-enhanced star formation, while the lower mass companion had its star formation suppressed. The enhancement of star formation in close pairs was reported by \cite{Ellison2008}. This picture is broadly consistent with the simulations presented by \cite{Moreno2015}, who showed that a close encounter between two galaxies will trigger the enhancement of star formation in a galaxy's centre.  It is unclear whether we are observing the same trend as \cite{Davies2015} and \cite{Ellison2008}, as they observed enhancement only in late-stage mergers, and we have a substantially smaller sample and fewer galaxy pairs at small separations. The trends that we have reported here apply to galaxies separated from their nearest neighbour by more than $30 \, \mathrm{kpc}$. 

Another possible explanation of our results comes from interpreting our data in the context of the discussion presented by \cite{Janowiecki2017}. In this work, the authors observe that in galaxies in groups with only two members, the central tends to have higher a HI content than galaxies in isolation at the same stellar mass. These HI-rich systems were observed to have higher sSFRs as well. From these observations, the authors suggest that gas-rich minor mergers or direct feeding of gas from the intergalactic medium may be more common in such environments. The acquisition of HI gas mass measurements for our sample would allow us to comment further on this point.

\subsection{Other metrics for interaction}

\subsubsection{Tidal Interactions}
We do not find any significant correlation between the estimated strength of the tidal interaction between galaxies and their star-forming properties. The apparent strength of the tidal force acting on a galaxy, calculated from Equation \ref{tid_pert_eqn}, influences neither the total measured specific star formation or the scale radius of star formation relative to the scale radius of the stellar light. Superficially this might seem to contradict results from simulations \citep{Hernquist89,Moreno2015} that suggest a central burst of star formation can occur in a galaxy after a close encounter with a companion if gas is present. However, our measurement of $P_{gc}$ is susceptible to systematic uncertainties imposed by projection effects when estimating the separation between two galaxies. The effect of projection will be to increase our estimate of the tidal interaction strength, and as such each measurement is at best an upper limit. A further shortcoming of this technique results from the fact that there is a delay between the time of closest approach for two systems and the time at which nuclear star formation will be triggered and is able to be measured. This delay, and the inability to distinguish between systems infalling towards an interaction and those that are moving away after an interaction, makes identifying the signatures of tidal interactions difficult with this technique. Therefore, we cannot rule out the possibility that tidal interactions cause quenching in high-mass galaxy groups or enhancement in low-mass groups.

\subsubsection{Close pair interactions}
Close-pair interactions have been reported to drive much of the environmental evolution of galaxies, including enhancing and suppressing their star formation \citep{LopezSanchez2008,LopezSanchez2009,Robotham14,Davies2015}. We have found no statistically significant link between the nearest neighbour distance and either $r_{50, \mathrm{H}\alpha}/r_{50,cont}$ or the sSFRs of galaxies in our sample. However, we note that our sample contains low numbers of galaxies at separations small enough to adequately test the predictions of these previous studies. A more comprehensive study of the distribution of star formation in close pairs of galaxies will be possible once the full SAMI survey has been completed and a larger sample of close pairs can be constructed.

\subsection{Comparison to other work}
Our results have built on the work presented in \cite{Schaefer17}, and we find general agreement with the trends presented therein. While our previous work compared the spatial extent of star formation to the fifth-nearest neighbour surface density environment measure, the use of galaxy group properties has provided a framework for a more physical understanding of the processes at play. In contrast to \cite{Schaefer17}, we find an anti-correlation between the scale-radius ratio and the stellar mass of the galaxies, but only find this in more massive group haloes. The quenching of galaxies with stellar mass greater than $10^{10} \, \mathrm{M_{\odot}}$ from the outside-in in dense environments is consistent with our previous findings, with the lack of this signature at lower stellar masses made more significant by our expanded sample.

These results from SAMI echo the findings from H$\alpha$ narrow-band imaging presented in \cite{Kulkarni15}. Kulkarni observed that galaxies with small scale-radius ratios lie below the star formation main sequence. The centrally-confined distribution of star formation in these galaxies, along with an observed flattening in the stellar light profiles in the outskirts of the galaxies, led them to conclude that a combination of ram pressure stripping and gravitational interactions are the primary mechanisms influencing group galaxies today. While with SAMI we are unable to investigate the outer stellar discs of our sample, we do find agreement in the star formation morphologies. A future study that combines the radial coverage of narrow-band imaging with the spectroscopic advantages of integral field surveys will yield important clues as to the relative impact of these two processes on shaping the galaxy populations of today. 

The radial distribution of star formation was also investigated by \cite{Spindler2018}, using data from the MaNGA survey. These authors found a very weak correlation between the star formation rate gradients in galaxies and the local environment density in comparison to \cite{Schaefer17}. This discrepancy can be reconciled by understanding the difference in the sample selection between the two studies. While \cite{Schaefer17} eliminated galaxies with AGN-like central spectra (in the same way as the Final-SF sample in the present work), \cite{Spindler2018} did not. The consequence of these choices is illustrated in Figure \ref{M_star_r50}. Comparing panels $c$ and $f$ shows that the inclusion of what \cite{Spindler2018} terms `centrally-suppressed' star-forming galaxies has the effect of washing out the more subtle signatures of environment quenching.

\cite{Bekki2014} produced hydrodynamical simulations of the ram pressure stripping of gas from galaxies in groups and clusters. With these simulations, they showed that the scale size of the star-forming discs of galaxies under ram pressure stripping can be reduced by a factor of two or more, depending on the halo mass of the group. The simulated galaxies under the influence of ram pressure stripping could have their star formation either enhanced or suppressed. For high-mass satellite galaxies ($M_{*}>10^{10} \, \mathrm{M_{\odot}}$) in our massive group sample, a reduction in the scale size of the star-forming disc is generally accompanied by a reduction in the total specific star-formation rate. The galaxies for which we do see a reduced scale radius ratio accompanied by an enhancement of the integrated star formation rate are the centrals of low-mass group halos and are unlikely to be undergoing ram pressure stripping and perhaps more likely to have undergone recent minor mergers or experienced fuelling from extragalactic gas \citep{Janowiecki2017}.

Analytical modeling by \cite{Hester2006} showed qualitatively similar results to \cite{Bekki2014}. In their work, they showed that the extent to which a galaxy is stripped of its gas by ram pressure depends on the mass distribution within a galaxy, and its trajectory through a cluster or group of a given mass. They showed that for galaxies moving through a group, it is possible for quenching to occur in just the outer parts of the disc. This provides a natural way of having ram pressure stripping quench star formation slowly or partially in the range of halo masses that we study here.

\section{Conclusion}\label{Conclusion}
We have used data from the SAMI Galaxy Survey to study the processes that suppress star formation in groups identified in the GAMA Galaxy Group Catalogue. The GAMA data provided several different metrics by which to quantify the environments of the galaxies in our sample. 

Our analysis shows a suppression of star formation in star--forming galaxies within high--mass groups ($M_G>10^{12.5}$\,M$_\odot$) at all galaxy stellar masses. This is hard to reconcile with models of instantaneous quenching.

We find that the concentration of star formation is correlated with galaxy stellar mass in high--mass groups.  Galaxies with stellar masses above approximately $10^{10} \, \mathrm{M_{\odot}}$ in high-mass groups are more likely to have concentrated star formation.  The fraction with concentrated star formation in the final SF sample is  $29^{+8}_{-7} \%$, compared to $4^{+3}_{-4} \%$ for similar galaxies that are not in groups. This central confinement of the star formation is also associated with a reduction in the total star formation rate, with no strong evidence for central enhancement of star-formation.  The concentration of star formation in these galaxies suggests a process such as ram--pressure stripping.  However, the location of these galaxies in the colour--mass plane, together with no separation between them and other galaxies in projected phase--space, suggests a relatively long timescale for this to occur. This mechanism appears only to act within $R_{200}$ of the centres of these massive groups.

In the same massive groups, galaxies with stellar masses less than $10^{10} \, \mathrm{M_{\odot}}$, show less evidence of centrally concentrated star formation.  This may infer a faster or more uniform quenching process, but one that still results in an overall reduction in the specific star formation rate of star--forming galaxies.

In lower mass groups ($M_G<10^{12.5}$\,M$_\odot$) we find no evidence of quenching.  Instead we find some evidence that centrally concentrated star formation is correlated with an increase in overall specific star formation rate.  This may be related to triggered star formation.

\section{Acknowledgements}

The SAMI Galaxy Survey is based on observations made at the Anglo-Australian Telescope. The Sydney-AAO Multi-object Integral field spectrograph (SAMI) was developed jointly by the University of Sydney and the Australian Astronomical Observatory. The SAMI input catalogue is based on data taken from the Sloan Digital Sky Survey, the GAMA Survey and the VST ATLAS Survey. The SAMI Galaxy Survey is supported by the Australian Research Council Centre of Excellence for All Sky Astrophysics in 3 Dimensions (ASTRO 3D), through project number CE170100013, the Australian Research Council Centre of Excellence for All-sky Astrophysics (CAASTRO), through project number CE110001020, and other participating institutions. The SAMI Galaxy Survey website is \url{http://sami-survey.org/}.

GAMA is a joint European-Australasian project based around a spectroscopic campaign using the Anglo-Australian Telescope. The GAMA website is \url{http://www.gama-survey.org/}.

MSO acknowledges the funding support from the Australian Research Council through a Future Fellowship (FT140100255). JTA acknowledges the award of a SIEF John Stocker Fellowship.
JvdS is funded under Bland-Hawthorn's ARC Laureate Fellowship (FL140100278). SB acknowledges the funding support from the Australian Research Council through a Future Fellowship (FT140101166). S.K.Y. acknowledges support from the Korean National Research Foundation (2017R1A2A1A05001116) and by the Yonsei University Future Leading Research Initiative (2015-22-0064). This study was performed under the umbrella of the joint collaboration between Yonsei University Observatory and the Korean Astronomy and Space Science Institute. CF gratefully acknowledges funding provided by the Australian Research Council's Discovery Projects (grants DP150104329 and DP170100603). Support for AMM is provided by NASA through Hubble Fellowship grant \#HST-HF2-51377 awarded by the Space Telescope Science Institute, which is operated by the Association of Universities for Research in Astronomy, Inc., for NASA, under contract NAS5-26555.

This research made use of Astropy, a community-developed core Python package for Astronomy \citep{Astropy}. We also used the Numpy and Scipy scientific python libraries.

\end{document}